# Complementarity of PALM and SOFI for super-resolution live cell imaging of focal adhesions

**Hendrik Deschout[1,5], Tomas Lukes[1,4,5], Azat Sharipov[1], Daniel Szlag[1], Lely Feletti[1], Wim Vandenberg[2], Peter Dedecker[2], Johan Hofkens[2], Marcel Leutenegger[3], Theo Lasser[1*], Aleksandra Radenovic[1*]**

[1]Laboratory of Nanoscale Biology & Laboratoire d'Optique Biomédicale, Ecole Polytechnique Fédérale de Lausanne, Lausanne, Switzerland
[2]Department of Chemistry, University of Leuven, Heverlee, Belgium
[3]Abteilung NanoBiophotonik, Max-Planck-Institut für biophysikalische Chemie, Göttingen, Germany
[4]Department of Radioelectronics, FEE, Czech Technical University in Prague, Prague, Czech Republic
[5]Equal contribution
*Corresponding author

## Abstract

**Live cell imaging of focal adhesions requires a sufficiently high temporal resolution, which remains a challenging task for super-resolution microscopy. We have addressed this important issue by combining photo-activated localization microscopy (PALM) with super-resolution optical fluctuation imaging (SOFI). Using simulations and fixed cell focal adhesion images, we investigated the complementarity between PALM and SOFI in terms of spatial and temporal resolution. This PALM-SOFI framework was used to image focal adhesions in living cells, while obtaining a temporal resolution below 10 s. We visualized the dynamics of focal adhesions, and revealed local mean velocities around 190 nm per minute. The complementarity of PALM and SOFI was assessed in detail with a methodology that integrates a quantitative resolution and signal-to-noise metric. This PALM and SOFI concept provides an enlarged quantitative imaging framework, allowing unprecedented functional exploration of focal adhesions through the estimation of molecular parameters such as the fluorophore density and the photo-activation and photo-switching rates**.

## 1. Introduction

It is essential for cells to adhere to the extracellular matrix for carrying out fundamental tasks such as migration, proliferation, and differentiation [1]. For all these processes, cell adhesion is essential. Focal adhesions rely on a concerted action of dense assemblies of hundreds of proteins [2] forming thin micron sized plaques close to the cell membrane [3]. These protein assemblies contain transmembrane receptors, such as integrins, binding to the extracellular matrix and recruiting other proteins inside the cytoplasm, such as talin and paxillin. This entails the formation of small structures in the order of 100 nm, which either disassemble after a few seconds, or mature into larger focal adhesions which remain stable typically for tens of minutes. This underlying maturation process requires an ongoing recruitment of additional proteins, such as vinculin or α-actinin, which may be linked to actin filaments. Overall, focal adhesions can thus be seen as the anchor points of the cell onto the extracellular matrix mediating interactions with the actin cytoskeleton. Most focal adhesion proteins have been identified. However, the observation of the spatial organization and dynamics of focal adhesions remains challenging.

Single molecule localization microscopy (SMLM), based on localizing sparse sets of activatable or switchable fluorescent molecules with a precision of tens of nanometers, is considered to be a method of choice for this endeavor [4] . In 2006, Betzig et al. [5] used photo-activated localization microscopy (PALM) to image the submicron patterns of vinculin in a fixed cell. However, focal adhesions are dynamic entities demanding fast live cell imaging. This has been further investigated by using PALM to image the dynamic behavior of paxillin [6], but elucidating the full cell adhesion process remains a challenging task for SMLM.

As shown in [6], SMLM trades temporal resolution for spatial super-resolution, since using less raw images for individual SMLM images means less available single molecule localizations. Several thousand raw images offer high spatial information of focal adhesions, but only a limited first glimpse into their dynamic behavior. These focal adhesions not only evolve over time, they can also undergo translational movement. The mean velocity of focal adhesions in stationary fibroblasts has been reported to be in the order of 100 nm per minute [7]. This translates into a temporal resolution well below one minute in order to capture the fundamental dynamic behavior while avoiding motion blur which would otherwise spoil the anticipated spatial resolution [6]. Although temporal resolutions in the order of seconds are possible using PALM [8], the SMLM method most often reported to achieve such a temporal resolution is (direct) stochastic optical reconstruction microscopy ((d)STORM) [9, 10]. However, delivery of (d)STORM dyes to intracellular targets remains difficult [11]. PALM is well suited for live cell imaging of focal adhesions since it uses genetically expressed fluorescent proteins known for being well tolerated in living cells.

PALM holds promise for obtaining information about the spatial composition and organization of proteins in focal adhesions. Indeed, assuming that each fluorescent protein is localized only once, their numbers would directly result in a fluorophore density map. However, fluorescent proteins are known to "blink", i.e. they can reversibly switch on and off for several times after being activated [12]. Blinking therefore results in an over-counting error. Several methods have been developed to account for this error, for instance by combining localizations that are clustered in space and time [13, 14] or by applying pair correlation analysis [15]. Under-counting errors can appear as well, not only by merging localizations of separate fluorophores in high density samples, but also due to incomplete maturation and limited detection efficiency [16].

In order to address the need for quantitative and time-lapse super-resolution imaging of focal adhesions, we enlarged the scope of SMLM by merging PALM with super-resolution optical fluctuation imaging (SOFI) [17] applied to the same raw image sequence. SOFI exploits the correlated response of neighboring image pixels based on a spatio-temporal cumulant analysis of image sequences [18]. This technique tolerates a significant overlap of single molecule images and relaxes the requirements on the activation or switching rates when compared with classical SMLM concepts. This allows one to use fluorescent molecules with a higher activation or switching rate [19], resulting in an improved temporal resolution [20]. However, there is a common belief that SOFI cannot attain the spatial resolution achievable by known SMLM methods. Additionally, balanced SOFI (bSOFI) can be used to determine the fluorophore on-time ratio, offering an estimation of the molecular density and molecular switching or activation rates [21].

In this paper, we investigated the complementarity of PALM and SOFI for imaging focal adhesions. By applying them both to the same dataset, we obtained a better insight in the true structure of focal adhesions and their molecular parameters. We have enhanced bSOFI and achieved a substantial increase in spatial resolution comparable to PALM. We also present a methodology for evaluating the super-resolution image quality, integrating a resolution and a signal-to-noise metric. We demonstrate our PALM-SOFI framework by imaging moving focal adhesions in a living cell.

## 2. Results

### 2.1 Widefield super-resolution metrics

In Abbe's theory, microscopy imaging is conceived as low pass filtering with a cut-off frequency at 2NA/λ (with λ the wavelength of light and NA the numerical aperture of the microscope objective). Abbe's analysis established the generally adopted resolution metric for classical microscopy as a pure instrument parameter independent of the object. SMLM goes beyond the "diffraction barrier" by exploiting to its best the precise localization of single fluorophores. Therefore, the final "SMLM-resolution" is the accumulated information of localized fluorescent markers and is de facto sample dependent.

In recent publications [22, 23], the concept of an optical resolution criterion was revisited with an extension to super-resolution imaging. However, as stated by Demmerle et al. [22], "resolution in single molecule imaging is especially challenging". There is a manifold of sample dependent and difficult to master parameters like labeling density, bleaching and the sample structure itself, which have a difficult to assess impact on resolution. In view of merging different imaging modalities like PALM and SOFI, the need for a general resolution and signal-to-noise (SNR) metric became mandatory.

An important step towards a resolution metric is the Fourier ring correlation (FRC) [24, 25]. Essential to this metric is the correlation of the Fourier transform of two SMLM images obtained from two stochastically independent halves of the original image sequence (see **Supplementary Note 1**). An extension of the FRC procedure applies also to SOFI, which we used for an objective assessment of PALM and SOFI. We imaged fixed mouse embryonic fibroblasts (MEFs) expressing paxillin labelled with mEos2 or psCFP2 (see **Methods**), and calculated the FRC metric as a function of the number of frames, as shown in **Figure 1**. In order to improve the spatial resolution of SOFI, we introduced a novel linearization procedure for bSOFI to achieve higher orders of the cumulant analysis (see **Supplementary Note 2**).

**Figure 1a-b** shows that the individual adhesion footprints are structured into a specific pattern. As the FRC calculation involves circular path summing in frequency space with a constant radius, the FRC metric is almost insensitive to variations of the spatial frequency content along different directions (see **Supplementary Figure 1**). In **Figure 1a-b**, such a difference can readily be noticed for the psCFP2 marked cell image, where focal adhesions and elongated structures indicative of paxillin organized along actin filaments [26] can be seen. We therefore implemented a sectorial FRC (sFRC) metric (see **Supplementary Note 1**) as already suggested by Nieuwenhuizen et al. [25]. This sFRC metric shows a more nuanced picture: the measured values are varying around the classic FRC for different sectors as shown in **Figure 1c** and **Supplementary Figure 2**, reflecting the orientation dependence of the resolution metric. The resolution capabilities of the imaging technique are best described by the sector with the lowest sFRC value, indicating that a spatial resolution around 100 nm was obtained. Interestingly, the sFRC values indicate that SOFI resolves psCFP2-expressing cells better than PALM, while the opposite was observed for mEos2 labelling, despite the latter fluorescent protein being well known for its blinking properties. We attribute these results to a difference in activation rate and emitter density, as indicated by the evolution of the number of localizations over time (see **Supplementary Figure 3**). The number of psCFP2 localizations is higher during the first several thousand frames, increasing the probability of overlapping psCFP2 images,

which poses more difficulties for PALM than for SOFI.

Besides the image resolution, the image SNR should be characterized as well. We performed a pixel-wise SNR estimation based on a statistical approach known as jackknife resampling [27]. The jackknife method generates N datasets of N-1 camera frames, i.e. each jackknife dataset is obtained by “cutting-out” just one single camera frame (see **Supplementary Note 3**). The variance on the individual pixel values originating from each of these datasets is considered as an uncertainty measure, yielding an SNR value per pixel. This general approach applies for PALM as well as for SOFI and has been used as an objective comparison of SNR for our PALM and SOFI cell images, as shown in **Figure 1d**. Except for a small number of frames (typically < 1000), PALM outperforms SOFI in terms of SNR. This is to be expected because the PALM images are reconstructed from fitted data.

In summary, our methodology for assessing the image quality integrates an objective evaluation of the resolution and the SNR for super-resolved images.

## 2.2 From spatial towards temporal resolution

Achieving a high temporal resolution in SMLM is truly a challenge. Bleaching, activation or switching rates, camera frame rates, and last but not least the minimum number of frames limit the achievable temporal resolution. As stated before, spatial super-resolution comes at an expense of temporal resolution. As we intend to image the dynamics of focal adhesions, we are in need of characterizing the difficult balance between lowering spatial super-resolution while enhancing temporal resolution. In order to objectively characterize the spatio-temporal resolution of both SOFI and PALM for a broad range of controlled conditions, we performed resolution measurements using simulated data. In an attempt to stay close to classical resolution measurement concepts, we designed a test target adoptedfrom charts used for modulation transfer function (MTF) analysis. The MTF allows one to extract the cut-off frequency and the visibility as a function of spatial frequency of an imaging system and is used as a metric for characterising optical imaging instruments [28]. Our MTF analysis provides a resolution standard for simulated data and a control for the sFRC resolution estimates in our high density conditions.

Our test target consists of progressively smaller bars randomly filled with point emitters at an a priori given density, providing an object of stochastically activated single emitters (see **Figure 2a** and **Supplementary Note 4**). To approximate the conditions of focal adhesions in a cell, we tested two labeling densities (i.e. 800 and 1200 molecules/μm$^2$). Our simulation takes into account the photophysics of mEos2 and psCFP2 and parameters of the microscope setup (see **Supplementary Note 4**). Based on this test target, we determined the visibility for PALM and SOFI beyond the cut-off frequency of classical widefield microscopy. From each simulated MTF, we extracted the cut-off frequency (see **Figure 2b** and **Supplementary Note 4**), resulting in a resolution measure related to the sFRC metric (see **Supplementary Figure 4**).

**Figure 2c-d** show the simulated cut-off frequency maps for PALM and SOFI based on the same test target, as a function of the number of frames and the number of photons per emitter per frame in an on-state (i.e. $I_{on}$). **Figure 2c** corresponds to 1200 molecules/μm$^2$ and the psCFP2 case, whereas **Figure 2d** corresponds to 800 molecules/μm$^2$ and the mEos2 case. The number of frames ranges from 500 to 20,000. At 20,000 frames, all molecules are detected and the structure of the test pattern is fully described. SOFI shows a slowly growing spatial resolution (i.e. an increase of cut-off frequency) with increasing $I_{on}$ and the number of frames. The PALM cut-off frequency grows faster, but only outperforms SOFI for a high number of frames (> 10,000 for the higher density case and > 5000 for the lower density case). Note that SOFI requires at least 500 frames before "super-resolution" can be achieved, while PALM needs even more frames (typically > 1000) and depends more strongly on the labeling density. For low frame numbers and low $I_{on}$, the number of localized emitters and the localization precision are too low for PALM to properly describe the test pattern, which results in low MTF values and corresponding low resolution. Assuming a typical camera frame rate of 100 Hz, **Figure 2e** shows the resolution sub-space where SOFI is dominant over PALM in terms of temporal/spatial resolution, and vice versa the sub-space where PALM outperforms SOFI. This indicates the parameter space where our PALM-SOFI imaging modality can be used for investigating the dynamics of focal adhesions as indicated in **Figure 2e**.

### 2.3 Live cell imaging

Imaging living cells requires a technique providing a sufficiently high temporal resolution and a compatibility with physiological conditions. Among the different SMLM methods, PALM meets the latter condition well, mainly due to genetically expressed fluorescent proteins acting as a label. However, the first condition is not perfectly met. PALM (like other SMLM techniques) makes the implicit assumption that the imaged structure stays stationary during the image acquisition, typically lasting for several minutes. Observing objects moving with a speed exceeding 10 nanometers per minute (i.e. the typical localization precision) is almost incompatible with this stationarity condition. Focal adhesions are known to move at rates of about 100 nm per minute, as mentioned before. Observing focal adhesion therefore demands PALM imaging cycles far below one minute, in order to avoid significant motion blur. The obvious way to increase the temporal resolution is to shorten the imaging cycle by acquiring less raw images. However, this entails a decrease in spatial resolution as less localizations are contributing. Many attempts have therefore been undertaken in SMLM to improve the temporal resolution, while maintaining a sufficient number of localizations [8, 29-31].

SOFI offers a large untapped potential for imaging living cells. Just like PALM, SOFI can be used with genetically expressed fluorescent proteins. However, it is also assumed that the sample is stationary during the acquisition of the raw images. This again asks for a tradeoff between spatial and temporal resolution, although SOFI images can be reconstructed with less images than required in PALM. When comparing SOFI and PALM, the latter technique is generally perceived as providing a higher spatial resolution. SOFI, on the other hand, is assumed to feature a higher temporal resolution, allowing faster imaging of moving structures, which has indeed been suggested by Geissbuehler et al. [20].

When attempting to increase both temporal and spatial resolution, a PALM-SOFI approach based on an identical raw image sequence appears as an interesting imaging modality. We imaged living MEFs expressing paxillin labelled with mEos2 and post-processed the data by both PALM and SOFI algorithms, as shown in **Figure 3a** and **Supplementary Video 1**. We obtained a temporal resolution of 10 s, while maintaining an average spatial resolution of 157 nm for SOFI, as determined by the sFRC metric (see **Supplementary Figure 5**). PALM at this temporal resolution resulted in an average spatial resolution of 145 nm. We determined the mean velocity of one of the focal adhesions, obtained from a kymograph based analysis [32, 33] (see **Figure 3b-c** and **Methods**). PALM and SOFI show similar trends, indicating that the focal adhesion moved with a mean velocity of 190 nm per minute. This mean velocity is in agreement with values reported and observed by others [7].

### 2.4 Quantitative imaging

Beyond qualitative imaging, SMLM methods such as PALM allow one to obtain quantitative molecular information, such as the number of localizations. This can be related to the number of fluorescent proteins. However, the relationship between both quantities is far from trivial, since most photoactivatable fluorescent proteins blink, i.e. they can reversibly go to a dark state. This may give rise to multiple localizations. Moreover, this blinking behavior depends on the illumination intensity and the molecular environment of the fluorescent proteins. Simply counting the localizations usually results in an overestimation of the number of fluorescent proteins. Hence, several methods to correct this over-counting error have been developed for PALM, often based on merging localizations that are sufficiently close in time and space to be considered originating from the same blinking fluorescent protein [13, 14]. As these methods require characterization of the blinking behavior, for instance through the calculation of the average time between two emission bursts, they indirectly allow one to probe the molecular environment of the emitters.

Focal adhesions are dense assemblies of proteins, making it challenging to avoid merging localizations of different fluorescent proteins, which would lead to an under-counting error. Therefore, we have adapted the merging criterion of an earlier published work [13] to account for higher densities. Instead of using a fixed distance threshold of 1 raw image pixel as merging criterion, we assumed a threshold based on a statistical measure, called the Hellinger distance, which allows one to account for the varying localization precision (see **Supplementary Note 5**). We applied this adapted method to our localization data (identical to those used for **Figure 1a-b**) of fixed MEFs expressing paxillin labeled with mEos2 or psCFP2, as shown in **Figure 4a,b,g,h**. The corrected localization number and the average time between two blinking events is shown as a function of different thresholds of the Hellinger distance, calculated for three areas with different emitter densities. We determined that a value of 0.9 was a good compromise (see **Supplementary Note 5**), but even around this value the number of localizations decreases with increasing threshold values for the densest areas (see **Figure 4d,j**). This indicates that the sample is too dense, which is corroborated by the average time between two blinking events being dependent on the area density (see **Figure 4e,k**).

SOFI is an interesting complement to PALM for quantitative imaging, since combining cumulant images of 2$^{nd}$, 3$^{rd}$, and 4$^{th}$ order enables to extract molecular parameters such as the on-time ratio, the molecular brightness, and the molecular density (see **Supplementary Note 2**) [21]. While PALM yields average values over the region of analysis, SOFI generates spatial maps of these parameters. Moreover, as SOFI is superior to PALM in imaging “crowded” environments, this method is of great interest for quantitative imaging of focal adhesions. We used SOFI to determine the on-time ratio and density map of the same localization data used for PALM (see **Figure 4c,f,i,l**). As opposed to PALM, SOFI performs well in high density areas. SOFI estimates the molecular parameters pixel-wise. This estimation is meaningless for areas which contain mostly background (SNR close to 1). Background areas therefore have to be removed by applying an intensity threshold or SNR based threshold. Since PALM is working well in these low density areas, this again demonstrates the usefulness of our PALM-SOFI approach.

## 3. Discussion

Our results indicate that PALM and SOFI are complementary techniques for the observation of focal adhesions in living cells. Such an imaging approach not only provides sufficient spatial resolution for their observation, it also grants access to their temporal dynamics. In view of the biological quest, we thoroughly investigated this imaging concept. Our simulations indicate a superior performance of SOFI when compared with PALM for low frame numbers (typically <5,000 frames), while PALM substantially outperforms SOFI for higher frame numbers. The onset of "super-resolution" based on SOFI demands typically 500 frames, while PALM requires at least 1000 frames. Our PALM-SOFI framework has been applied to the same raw image sequences therefore opens the door for assessing the dynamics of "not too fast" biological processes in the order of 100 nm per minute.

Using both PALM and SOFI, we could image focal adhesions with a resolution better than 100 nm in fixed cells, whereas in living cells a resolution < 150 nm was obtained, requiring less than 1000 raw images. These live cell images were recorded at a frame rate of 100 Hz, which translates into a temporal resolution below 10 s. Such a temporal resolution is required to resolve the dynamics of the focal adhesions in more detail, as we observed "focal adhesion velocities" around 190 nm per minute.

Besides resolution, we also characterized the SNR for our PALM-SOFI framework. In general, PALM shows the highest SNR, up to 25 dB for large frame numbers for fixed cell images. Only for small frame numbers (typically < 500) SOFI showed a superior SNR. We attribute this difference to the different nature of PALM and SOFI (i.e. image rendered based on localized emitters versus correlations of intensity fluctuations). Considering this difference, the ramp up towards the SNR plateau seems to be more important for our data than a comparison of absolute SNR values (see **Supplementary Figure 6**). The steeper onset of SNR is in favor of PALM whereas for SOFI the SNR plateau is reached at a lower frame number.

We used a generalized resolution metric named sFRC (adapted from the classical FRC metric) and a SNR metric based on statistical resampling for assessing the performance gain of the PALM-SOFI framework. Our simulations show that the sFRC metric is in agreement with the cut-off frequencies obtained from our MTF analysis (see **Supplementary Figure 4**). Under the tested conditions corresponding to focal adhesions, the (s)FRC values are slightly higher than expected. We attribute this to the fixed threshold used in the calculation of the (s)FRC metric. We would like to note that the sFRC by definition requires images with a rich spatial frequency content. When, for instance, a sparse structure in the presence of mostly background is imaged, the sFRC value is unreliable, and this metric is useful for qualitative comparison only.

Depending on the fluorophore properties, either PALM or SOFI yielded a better resolution. Using mEos2, PALM performed better, while SOFI outperformed PALM for psCFP2. We hypothesize that this is caused by a difference in activation rate, combined with a difference in fluorophore density. For mEos2, the localization number per frame was low and constant, which is in favor of PALM. psCFP2, on the other hand, showed a higher number of localizations during the beginning of the image acquisition, resulting in a better resolution for SOFI. This points to the interesting fact that difficult PALM data, caused by a "crowded" environment, can still be evaluated by SOFI.

Further benefits of this PALM-SOFI complementarity have been demonstrated by applying quantitative analysis on our focal adhesion data. PALM was shown to give reliable estimates of the blinking corrected localization numbers and the off-time between blinks in low density areas of the cell sample. SOFI, on the other hand, was able to extract on-time ratios and number densities in high density regions.

## 4. Methods

### 4.1 Microscope set-up

Fixed cell imaging was carried out on a custom built microscope [34]. Three continuous wave laser sources were used for excitation/activation: a 50 mW 405 nm laser (Cube, Coherent), a 100 mW 488 nm laser (Sapphire, Coherent), a 100 mW 561 nm laser (Excelsior, Spectra Physics). The 488 nm and 561 nm lasers were combined using a dichroic mirror (T495lpxr, Chroma) and sent through an acousto-optic tunable filter (AOTFnC-VIS-TN, AA Opto Electronic). Both lasers were combined with the 405 nm laser using a dichroic mirror (405 nm laser BrightLine, Semrock). The three lasers were focused by a lens into the back focal plane of the objective mounted on an inverted optical microscope (IX71, Olympus). We used a 100× objective (UApo N 100×, Olympus) with a numerical aperture of 1.49 configured for total internal reflection fluorescence microscopy (TIRF). The fluorescence light collected by the objective was filtered to suppress the residual illumination light using a combination of a dichroic mirror (493/574 nm BrightLine, Semrock) and an emission filter (405/488/568 nm StopLine, Semrock). The fluorescence light was detected by an EMCCD camera (iXon DU-897, Andor). The back-projected pixel size was 105 nm. An adaptive optics system (Micao 3D-SR, Imagine Optics) and an optical system (DV2, Photometrics) equipped with a dichroic mirror (617/73 nm BrightLine, Semrock) were placed in front of the EMCCD camera. The Micao 3D-SR system was used to compensate for aberrations and the DV2 system was used to split the fluorescence light into a green and red color channel that were each sent to a separate half of the camera chip.

Live cell imaging was carried out on a custom built microscope equipped with a temperature and $CO_2$ controlled incubator for live cell imaging [20]. Three continuous wave laser sources were used for excitation/activation: a 120 mW 405 nm laser (iBeam smart, Toptica), a 200 mW 488 nm laser (iBeam smart, Toptica), a 800 mW 532 nm laser (MLL-FN-532, Roithner Lasertechnik). The 488 nm and 532 nm lasers were combined using a dichroic mirror (T495LP, Chroma), and both lasers were combined with the 405 nm laser using a dichroic mirror (T425LPXR, Chroma). All three lasers were focused by a lens into the back focal plane of the objective. We used a 60× objective (Apo N 60×, Olympus) with a numerical aperture of 1.49 that allows for TIRF illumination. The fluorescence light collected by the objective was filtered to suppress the residual illumination light using a combination of a dichroic mirror (Z488/532/633RPC, Chroma) and an emission filter (ZET405/488/532/640m, Chroma). The fluorescence light was detected by an EMCCD camera (iXon DU-897, Andor). The back projected pixel size was 96 nm.

### 4.2 Sample preparation

The mouse embryonic fibroblasts (MEFs) were grown in DMEM supplemented with 10% fetal bovine serum, 1% penicillin-streptomycin, 1% non-essential amino acids and 1% glutamine, at 37 °C with 5% $CO_2$. The cells were transfected by electroporation (Neon Transfection System, Invitrogen), which was performed on 600,000-800,000 cells using 1 pulse of 1350 V lasting for 35 ms. The amount of DNA used for the transfection was 2 µg for the mEos2-paxillin-22 vector and 5 µg for the psCFP2-paxillin-22 vector.

For fixed cell experiments, a 25 mm diameter microscope cover slip (#1.5 Micro Coverglass, Electron Microscopy Sciences) was prepared by first cleaning with an oxygen plasma for 5 minutes and then incubated with PBS containing 50 µg/ml fibronectin (Bovine Plasma Fibronectin, Invitrogen) for 30 minutes at 37 °C.

To remove the excess of fibronectin, the cover slip was washed with PBS. The transfected cells were seeded on the cover slip and grown in DMEM supplemented with 10% fetal bovine serum, 1% non-essential amino acids and 1% glutamine, at 37 °C with 5% $CO_2$. The cells were washed with PBS around 24 h after transfection, and then incubated in PBS with 4% paraformaldehyde at 37 °C for 30 minutes. After removing the fixative, the cells were again washed with PBS, and the cover slip was placed into a custom made holder.

For live cell imaging, the transfected cells were seeded in a chambered cover slip system (Lab-Tek II Chambered Coverglass System, Thermo Scientific) and grown in DMEM supplemented with 10% fetal bovine serum, 1% non-essential amino acids and 1% glutamine, at 37 °C with 5% $CO_2$. Finally, the cells were washed with PBS around 24 h after transfection.

### 4.3 Imaging procedure

Fixed cells were imaged in PBS at room temperature. Prior to imaging, 100 nm gold nanospheres (C-AU-0.100, Corpuscular) have been added to the sample for lateral drift monitoring. Axial drift correction was ensured by a nanometer positioning stage (Nano-Drive, Mad City Labs) driven by an optical feedback system [34]. Excitation of mEos2 was done at 561 nm with ~15 mW power (as measured in the back focal plane of the objective). Imaging of psCFP2 was performed using 488 nm laser light with ~15 mW power. Both fluorescent labels were gently activated by 405 nm laser light with ~5 µW power. The gain of the EMCCD camera was set at 100 and the exposure time to 50 ms. For each experiment at least 20,000 camera frames were recorded.

The live cells were imaged in DMEM with low fluorescence background (FluoroBrite DMEM, Thermo Scientific) at 37 °C with 5% $CO_2$. Prior to imaging, 100 nm gold nanospheres (C-AU-0.100, Corpuscular) were added to the sample for lateral drift correction. mEos2 was excited at 532 nm with ~8.5 mW power and activated by 405 nm laser light with ~0.6 mW power. The gain of the EMCCD camera was set at 150 and the exposure time to 10 ms. For each experiment at least 8,000 camera frames were recorded.

### 4.4 PALM data analysis

The recorded images were analyzed by a custom written algorithm (Matlab, The Mathworks) that was adapted from an algorithm that was published elsewhere [5]. First, peaks were identified in each camera frame by filtering the frames and applying an intensity threshold. Only peaks with an intensity of at least 4 times the background were considered to be fluorescent labels or gold nanospheres. Subsequently, the peaks were fitted by maximum likelihood estimation of a 2D Gaussian distribution [35]. Drift was corrected in each frame by subtracting the average position of the gold nanospheres from the positions of the fluorescent labels that were localized in that frame. The theoretical localization precision for each fluorescent label was obtained from the Cramér-Rao lower bound of the maximum likelihood procedure [36]. This value was multiplied with the square root of 2 in order to account for the degradation of the localization precision caused by the electron multiplication process in the EMCCD camera [35]. The PALM image was generated either as a 2D localization number histogram or by plotting a 2D Gaussian centered on each fitted position with a standard deviation equal to the corresponding localization precision. Only positions with a localization precision between 1 and 50 nm were plotted.

### 4.5 SOFI data analysis

For the SOFI calculation, we adapted and enhanced the bSOFI algorithm [21] (see **Supplementary Note 2**). The whole input image sequence was divided into subsequences of 500 frames each. The subsequences were chosen sufficiently short to minimize the influence of photobleaching. As SOFI assumes the sample to be stationary over the investigated image subsequence, drift has to be corrected prior to the bSOFI processing. Tracking the positions of the gold nanospheres provides translational motion vectors in between consecutive frames. Drift was then corrected by registering the frames with sub-pixel precision using bilinear interpolation. The linearization step of the bSOFI algorithm was replaced by an adaptive linearization which takes into account blinking properties of the sample and thus enablesmore effective use of the available dynamic range and SNR for high order SOFI analysis.

### 4.6 Simulations

For each fluorophore, a time trace was modelled, describing the number of photons emitted by a given fluorophore over time. The simulation assumed photokinetics typical for fluorescent proteins in PALM experiments (see **Supplementary Note 4**). The intensity of a pixel at a certain time point was given by an integration of brightness from all fluorophores with a PSF that extends to that pixel at that time point. The number of photo-electrons was simulated by a Poisson distributed random number with an average value equal to the pixel value multiplied by the detection efficiency and added to the thermal noise (dark current). Gain noise and read-out noise were modelled as additive Gaussian noise. The parameters of the optical system and camera used for simulations matched the properties of the microscope set-up. We tested two labelling densities: 800 and 1200 molecules/$\mu m^2$, leading to two different scenarios (see **Figure 2c-d**). For each scenario, the number of photons per emitter per frame (i.e. $I_{on}$) varied from 50 to 400 and the number of frames ranged from 500 to 20,000. In total, we generated and analysed 60 image stacks. Each image sequences was processed by a SMLM and a bSOFI algorithm. For SMLM processing, we used the FALCON algorithm [37] with the settings tuned for high densities. Using the bSOFI algorithm, images of $2^{nd}$ to $6^{th}$ order were calculated. The cut-off frequency was measured for every bSOFI order. With increasing order of the bSOFI analysis, the resolution increases, but the image SNR decreases which limits the highest achievable resolution. The output SOFI cut-off frequencies shown in **Figure 2c-d** represent the highest cut-off frequency achieved from the measured orders of the bSOFI analysis.

### 4.7 Measuring the cut-off frequency

An average line profile was calculated from each simulated super-resolved output image. The one-dimensional MTF (see **Supplementary Note 4**) was calculated as the modulus of the discrete Fourier transform of the average line profile. Each MTF curve was smoothed by a moving average filter with a window length equal to three in order to suppress fluctuations and provide more robust estimate of the cut-off frequency. To eliminate small non-zero MTF values which are caused mostly by noise and do not contain relevant information, we subtracted a constant 0.5 from each MTF curve prior to normalization. Each MTF curve was normalized using the MTF corresponding to the 20,000 frames test case as a reference. The cut-off frequency is the spatial frequency where the normalized MTF curve falls below a threshold (i.e. a small positive constant close to zero). The threshold was determined as the value of the widefield MTF which occurs at the theoretical cut-off frequency of a noiseless diffraction-limited imaging system given by Abbe's resolution limit.

### 4.8 sFRC calculation

We used the sFRC metric for analyzing the images shown in **Figure 1a-b**. The full raw image sequence (20,000 frames) was split into 40 subsequences of 500 frames each (see **Supplementary Note 1**). For bSOFI, images up to the $6^{th}$ order were calculated for each subsequence. These images were split into two groups and averaged within each group to generate two SOFI images. The splitting procedure is described in **Supplementary Note 1**. For PALM, the localizations corresponding to the selected 500 frame subsequences were pooled and used to render two independent PALM images as 2D histograms with a pixel size that is approximately 1/6 of the real pixel size, matching the $6^{th}$ order bSOFI pixel size. The recombination into two independent PALM/SOFI images was done according to the procedure described in the **Supplementary Note 1**. In order to observe the evolution of the sFRC with increasing number of frames, the calculation was repeated using an increasing amount of frames, going from 1000 to 20,000 frames with an increment of 1000 frames in each step. The sFRC was calculated in separate sectors with an angular extent of $\pi/12$. The results for all sectors are shown in **Supplementary Figure 2**. Two selected sectors are shown in **Figure 1c**.

### 4.9 SNR calculation

We calculated the pixel-wise SNR using a statistical approach, i.e. jackknife resampling (see **Supplementary Note 3**) on the data shown in **Figure 1a-b**. For an objective comparison, PALM images were rendered as histograms with a pixel size of 105 nm (i.e. the pixel size in the raw images) and SOFI images were binned on an equal pixel size prior to the SNR estimation. In order to observe the evolution of the SNR throughout the raw image sequence (20,000 frames), the calculation was repeated for an increasing number of frames, starting with 1000 frames and adding the next 1000 frames in each step. The SNR values as function of the number of frames are shown in **Figure 1d**.

### 4.10 Kymograph based analysis

The kymograph shown in **Figure 3c** along the line indicated in **Figure 3b** was obtained using ImageJ, (National Institutes of Health). For each time point, the center position of the focal adhesion was calculated as the center of gravity along the corresponding line in the kymograph, with the PALM/SOFI pixel values as weights. The focal adhesion mean velocity was determined as the slope of a linear fit to these center positions as a function of the time points. This procedure was repeated for 4 other lines parallel to the one shown in **Figure 3b**. The reported focal adhesion mean velocity is the average and the error bar represents the corresponding standard deviation. The direction of the kymograph was chosen as the direction of the focal adhesion mean velocity, which was determined by applying the above procedure to the x- and y-direction separately (see **Supplementary Figure 7**).

## Acknowledgements

The MEFs were kindly provided by Dr. Luca Scorrano. The mEos2-paxillin-22 and psCFP2-paxillin-22 vectors were kindly provided by Dr. Michael Davidson. H.D. and A.R. acknowledge the support of the Max Planck-EPFL Center for Molecular Nanoscience and Technology.

## Author contributions

H.D., T.Lu., T.La. and A.R. conceived the study. T.Lu., T.La. and D.S. developed the MTF analysis. T.Lu. performed the simulations. H.D. and L.F. prepared the samples. H.D. performed the fixed cell experiments, H.D. and A.S. performed the live cell experiments. T.Lu. developed the enhanced bSOFI algorithm. T.Lu. and H.D. analyzed the data. W.V., M.L. developed the Jackknife code. D.S., W.V., P.D., J.H. and M.L. provided research advice. H.D., T.Lu., T.La. and A.R. wrote the paper. All authors reviewed and approved the manuscript.

## References

1. Geiger, B., J.P. Spatz, and A.D. Bershadsky, *Environmental sensing through focal adhesions.* Nature Reviews Molecular Cell Biology, 2009. **10**(1): p. 21-33.
2. Zaidel-Bar, R., et al., *Functional atlas of the integrin adhesome.* Nature Cell Biology, 2007. **9**(8): p. 858-868.
3. Franz, C.M. and D.J. Muller, *Analyzing focal adhesion structure by atomic force microscopy.* Journal of Cell Science, 2005. **118**(22): p. 5315-5323.
4. Tabarin, T., et al., *Insights into Adhesion Biology Using Single-Molecule Localization Microscopy.* Chemphyschem, 2014. **15**(4): p. 606-618.
5. Betzig, E., et al., *Imaging intracellular fluorescent proteins at nanometer resolution.* Science, 2006. **313**(5793): p. 1642-1645.
6. Shroff, H., et al., *Live-cell photoactivated localization microscopy of nanoscale adhesion dynamics.* Nature Methods, 2008. **5**(5): p. 417-423.
7. Smilenov, L.B., et al., *Focal adhesion motility revealed in stationary fibroblasts.* Science, 1999. **286**(5442): p. 1172-1174.
8. Huang, F., et al., *Video-rate nanoscopy using sCMOS camera-specific single-molecule localization algorithms.* Nature Methods, 2013. **10**(7): p. 653-658.
9. Carlini, L. and S. Manley, *Live Intracellular Super-Resolution Imaging Using Site-Specific Stains.* Acs Chemical Biology, 2013. **8**(12): p. 2643-2648.
10. Shim, S.H., et al., *Super-resolution fluorescence imaging of organelles in live cells with photoswitchable membrane probes.* Proceedings of the National Academy of Sciences of the United States of America, 2012. **109**(35): p. 13978-13983.
11. Hennig, S., et al., *Instant Live-Cell Super-Resolution Imaging of Cellular Structures by Nanoinjection of Fluorescent Probes.* Nano Letters, 2015. **15**(2): p. 1374-1381.
12. Annibale, P., et al., *Identification of clustering artifacts in photoactivated localization microscopy.* Nature Methods, 2011. **8**(7): p. 527-528.
13. Annibale, P., et al., *Quantitative Photo Activated Localization Microscopy: Unraveling the Effects of Photoblinking.* Plos One, 2011. **6**(7).
14. Lee, S.H., et al., *Counting single photoactivatable fluorescent molecules by photoactivated localization microscopy (PALM).* Proceedings of the National Academy of Sciences of the United States of America, 2012. **109**(43): p. 17436-17441.
15. Sengupta, P. and J. Lippincott-Schwartz, *Quantitative analysis of photoactivated localization microscopy (PALM) datasets using pair-correlation analysis.* Bioessays, 2012. **34**(5): p. 396-405.
16. Durisic, N., et al., *Single-molecule evaluation of fluorescent protein photoactivation efficiency using an in vivo nanotemplate.* Nature Methods, 2014. **11**(2): p. 156–162.
17. Vandenberg, W., et al., *Diffraction-unlimited imaging: from pretty pictures to hard numbers.* Cell and Tissue Research, 2015. **360**(1): p. 151-178.
18. Dertinger, T., et al., *Fast, background-free, 3D super-resolution optical fluctuation imaging (SOFI).* Proceedings of the National Academy of Sciences of the United States of America, 2009. **106**(52): p. 22287-22292.
19. Geissbuehler, S., C. Dellagiacoma, and T. Lasser, *Comparison between SOFI and STORM.* Biomedical Optics Express, 2011. **2**(3): p. 408-420.
20. Geissbuehler, S., et al., *Live-cell multiplane three-dimensional super-resolution optical fluctuation imaging.* Nature Communications, 2014. **5**.

21. Geissbuehler, S., et al., *Mapping molecular statistics with balanced super-resolution optical fluctuation imaging (bSOFI).* Optical Nanoscopy, 2012. **1**(4).
22. Demmerle, J., et al., *Assessing resolution in super-resolution imaging.* Methods, 2015. **88**: p. 3-10.
23. Li, D., et al., *Extended-resolution structured illumination imaging of endocytic and cytoskeletal dynamics.* Science, 2015. **349**(6251).
24. Banterle, N., et al., *Fourier ring correlation as a resolution criterion for super-resolution microscopy.* Journal of Structural Biology, 2013. **183**(3): p. 363-367.
25. Nieuwenhuizen, R.P.J., et al., *Measuring image resolution in optical nanoscopy.* Nature Methods, 2013. **10**(6): p. 557-562.
26. Shroff, H., et al., *Dual-color superresolution imaging of genetically expressed probes within individual adhesion complexes.* Proceedings of the National Academy of Sciences of the United States of America, 2007. **104**(51): p. 20308-20313.
27. Vandenberg, W., et al., *Model-free uncertainty estimation in stochastical optical fluctuation imaging (SOFI) leads to a doubled temporal resolution.* Biomedical Optics Express, 2016. **7**(2): p. 467-480.
28. *Handbook of optical systems, volume 3, aberration theory and correction of optical systems*, ed. H. Gross. 2005, New York: Wiley.
29. Cox, S., et al., *Bayesian localization microscopy reveals nanoscale podosome dynamics.* Nature Methods, 2012. **9**(2): p. 195-200.
30. Ovesny, M., et al., *High density 3D localization microscopy using sparse support recovery.* Optics Express, 2014. **22**(25): p. 31263-31276.
31. Wang, Y., et al., *PALMER: a method capable of parallel localization of multiple emitters for high-density localization microscopy.* Optics Express, 2012. **20**(14): p. 16039-16049.
32. Rossier, O., et al., *Integrins beta(1) and beta(3) exhibit distinct dynamic nanoscale organizations inside focal adhesions.* Nature Cell Biology, 2012. **14**(10): p. 1057–1067.
33. Stehbens, S.J., et al., *CLASPs link focal-adhesion-associated microtubule capture to localized exocytosis and adhesion site turnover.* Nature Cell Biology, 2014. **16**(6): p. 558–570.
34. Annibale, P., et al., *Identification of the factors affecting co-localization precision for quantitative multicolor localization microscopy.* Optical Nanoscopy, 2012. **1**(9).
35. Mortensen, K.I., et al., *Optimized localization analysis for single-molecule tracking and super-resolution microscopy.* Nature Methods, 2010. **7**(5): p. 377-381.
36. Ober, R.J., S. Ram, and E.S. Ward, *Localization accuracy in single-molecule microscopy.* Biophysical Journal, 2004. **86**(2): p. 1185-1200.
37. Min, J.H., et al., *FALCON: fast and unbiased reconstruction of high-density super-resolution microscopy data.* Scientific Reports, 2014. **4**.

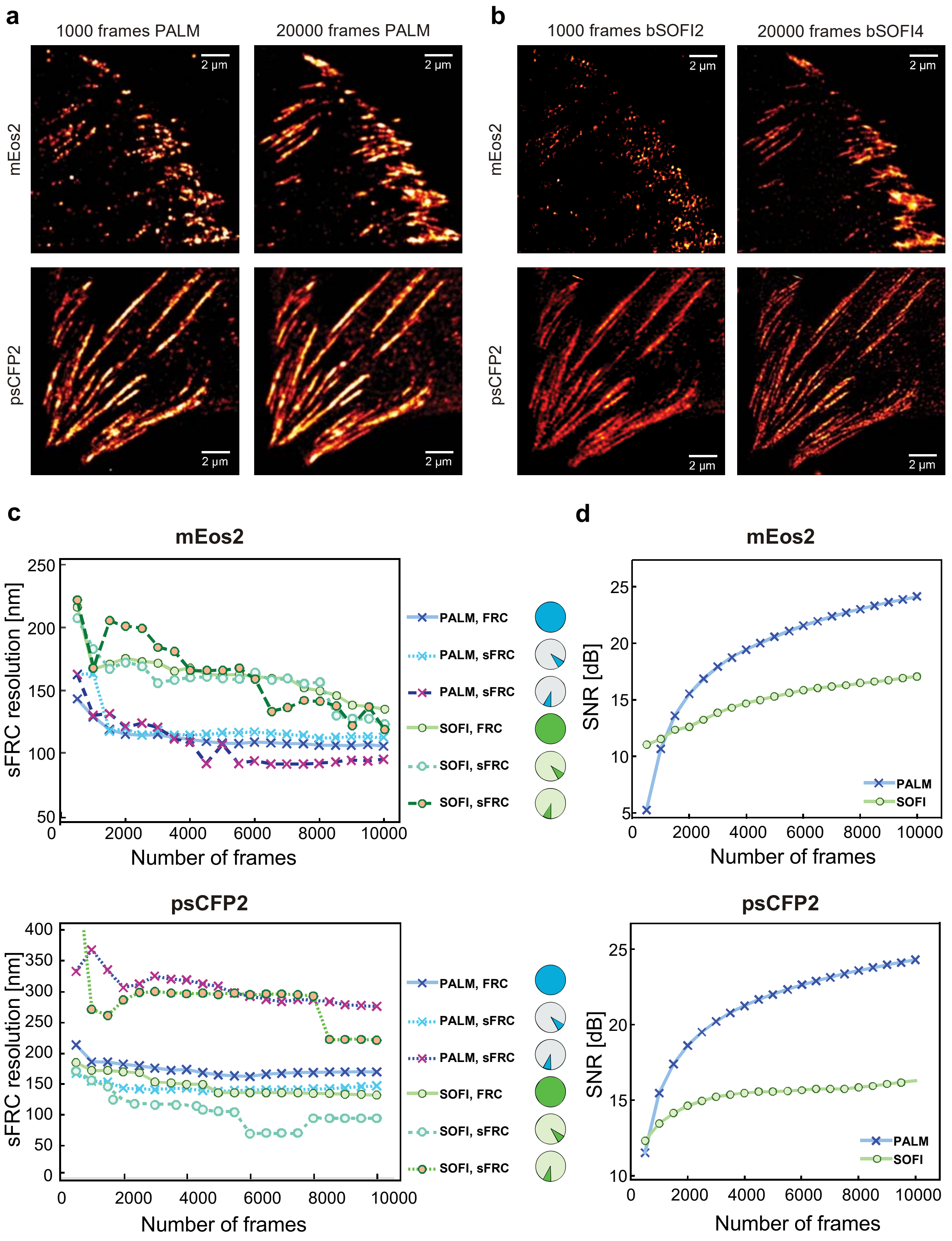

**Figure 1:** Objective image quality assessment integrating a resolution metric and a SNR metric applied to PALM and SOFI images. (a) PALM images of fixed MEFs expressing paxillin labeled with mEos2 or psCFP2, obtained from a full raw image sequence (20,000 frames) and the first 1000 frames. (b) SOFI images obtained from the same raw image sequences as in (a). (c) Resolution (sFRC) metric for SOFI and PALM as a function of the number of frames, obtained from subsequences of the same raw image sequences as in (a-b). The circles indicated the sector used for the sFRC calculation, the sector with the lowest sFRC values provides the best description of the resolution. Note that the sFRC requires to split the number of frames in two halves to create two images. Therefore, 20,000 input frames allows one to calculate the sFRC corresponding to a super-resolved image reconstructed from 10,000 frames. (d) SNR metric for SOFI and PALM as a function of the number of frames, obtained from subsequences of the same raw image sequences as in (a-b).

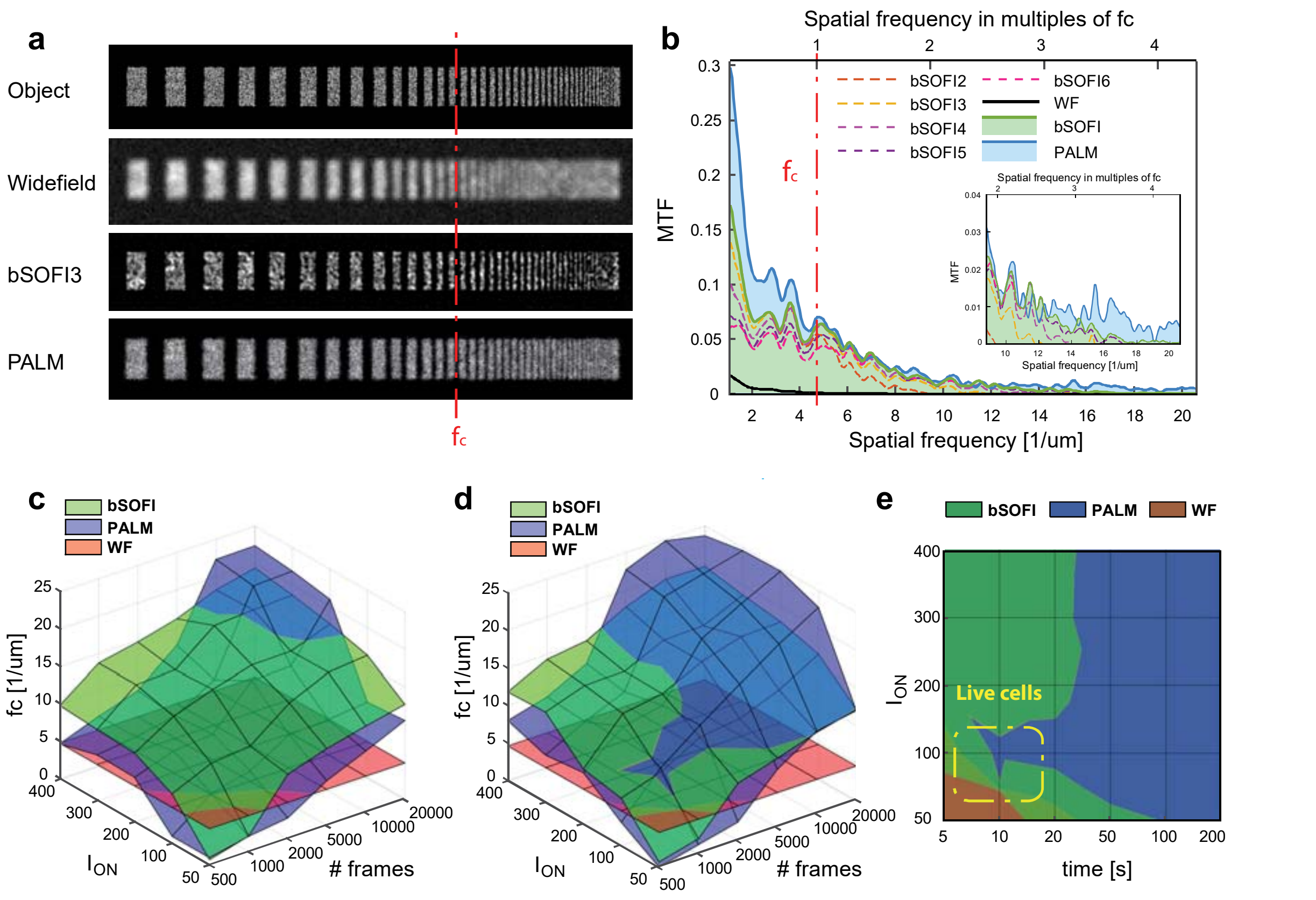

**Figure 2:** MTF analysis on simulated PALM and SOFI images. (a) MTF test target consisting of single emitters randomly placed inside progressively thinner bars, together with resulting widefield, PALM, and 3rd order SOFI image ($I_{on}$ = 100 photons and 20,000 frames) The red line indicates the cut-off frequency for widefield imaging. (b) MTF calculated from the simulated SOFI and PALM images in (a). (c-d) Cut-off frequencies for PALM and SOFI as a function of $I_{on}$ and the number of frames, with an emitter density of (c) 1200 #/µm$^2$ and (d) 800 #/µm$^2$. (e) Two-dimensional projection of the chart in (d). The timescale assumes a frame rate equal to 100 Hz which corresponds to the frame rate used for our live cell measurements.

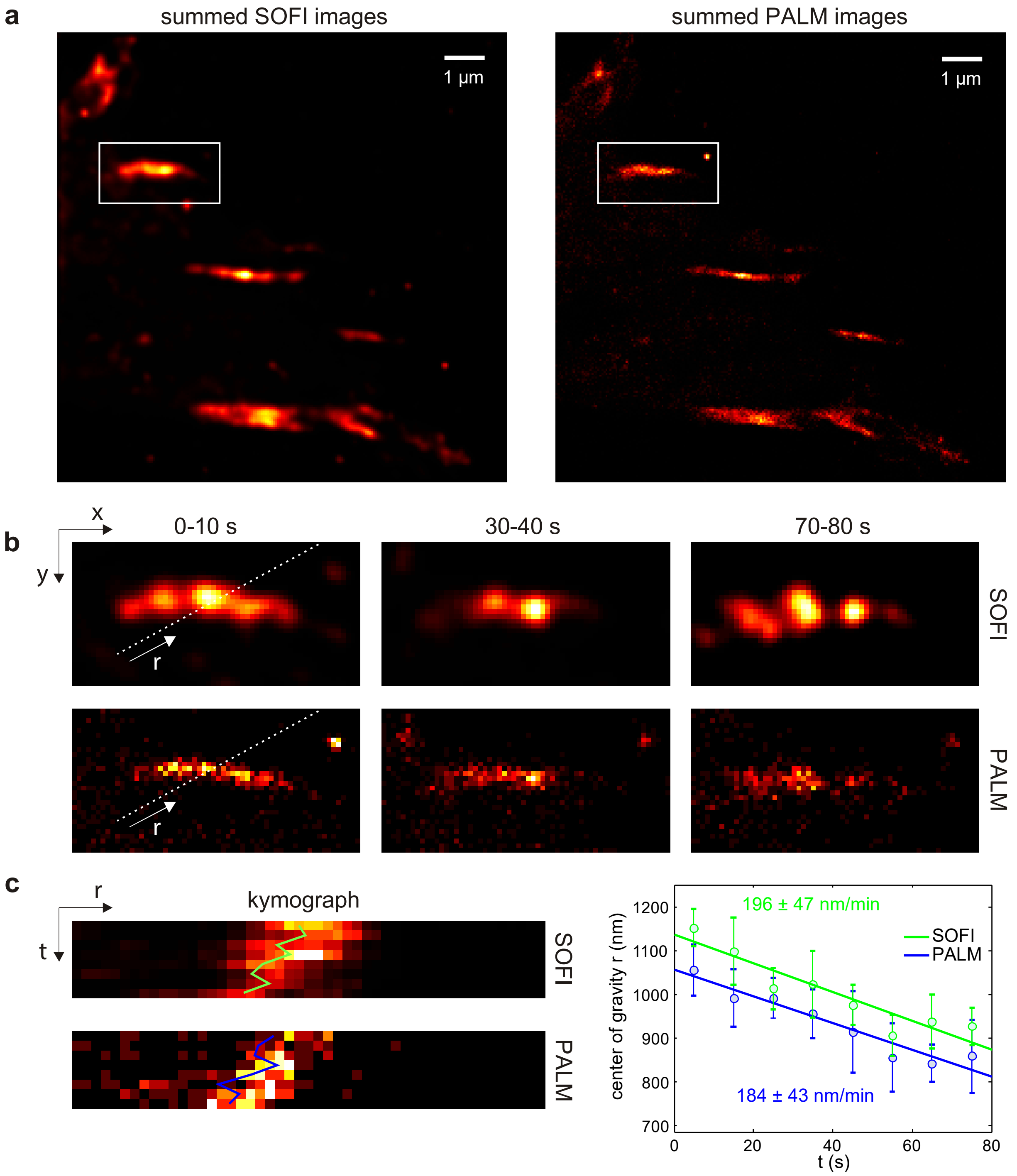

**Figure 3:** Live cell imaging with PALM and SOFI. (a) Sum of 8 PALM and SOFI images of a living MEF expressing paxillin labeled with mEos2. Each image is reconstructed from 1000 camera frames with 10 ms exposure time, resulting in a 10 s temporal resolution. (b) Region of interest indicated in (a) showing a focal adhesion at different time points. (c) Kymographs along the direction of motion as indicated by the line indicated in (b). The focal adhesion mean velocity is determined by a linear fit to the center

position determined from the kymograph as a function of time. The procedure was repeated 5 times for parallel kymographs.

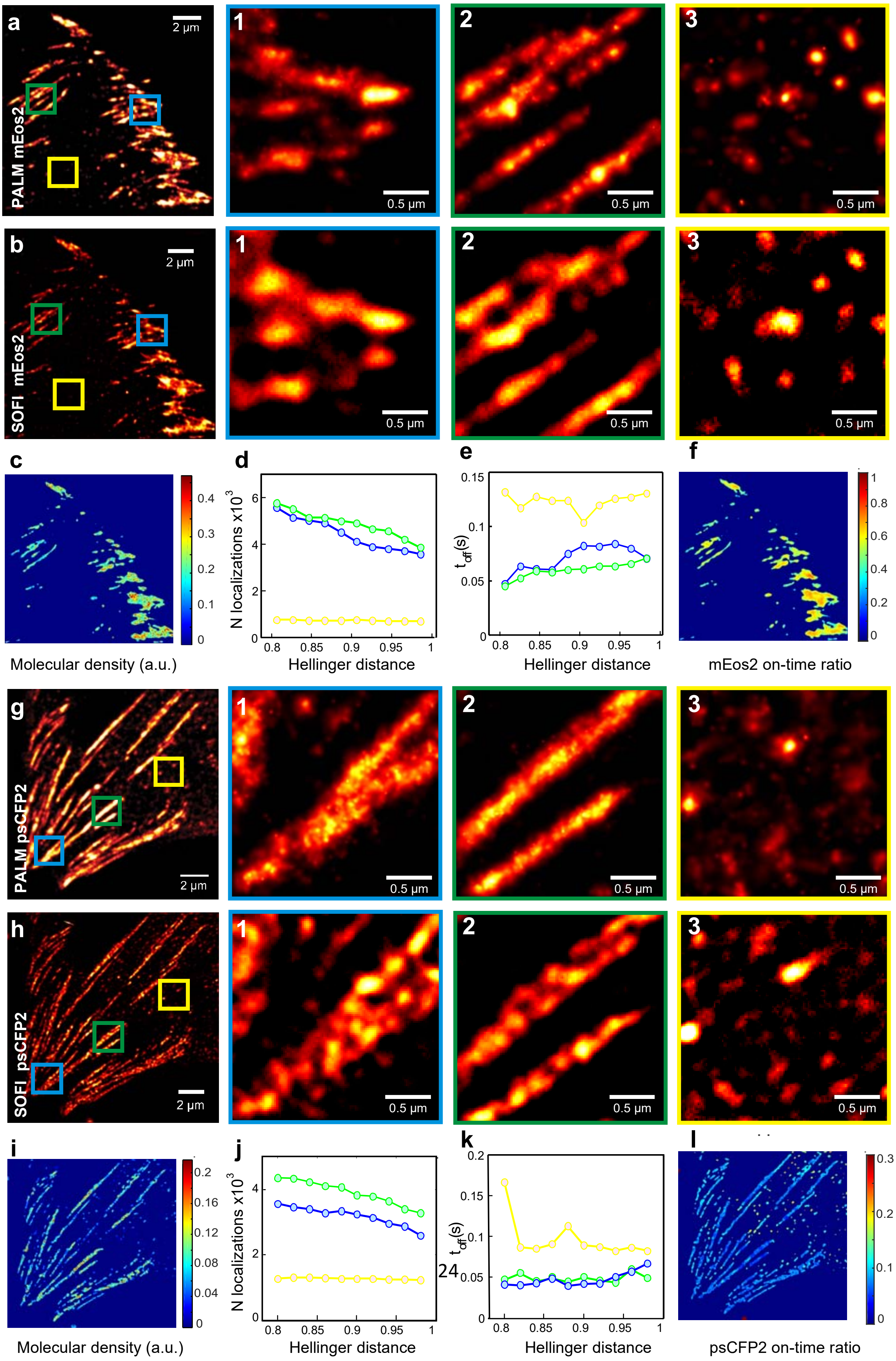

**Figure 4:** Quantitative imaging with PALM and SOFI. (a-b) PALM and SOFI images of a fixed MEF expressing paxillin labeled with mEos2. Panels 1-3 are corresponding zoom-ins for PALM or SOFI images. (c,f) Blinking events in PALM data can be detected by merging localizations that are sufficiently close in space and time. This analysis yields the blink corrected number of localizations N and the corresponding average off-time $t_{off}$ between blinks, shown as a function of the distance threshold for merging localizations. (d,e) SOFI analysis yields a fluorophore density and on-time ratio map. (g-h) PALM and SOFI images of a fixed MEF expressing paxillin labelled with psCFP2. Panels 1-3 are corresponding zoom-ins for PALM or SOFI images. (i-l) The same quantitative analysis as shown in (c-f).

# Supplementary Material: Complementarity of PALM and SOFI for super-resolution live cell imaging of focal adhesions

This paper is devoted to focal adhesions and their dynamics. In view of this important biological phenomenon, we investigated the spatio-temporal resolution of photo-activated localization microscopy (PALM) and super-resolution optical fluctuation imaging (SOFI) when imaging focal adhesions in fixed and living cells, by applying both methods on the same image sequence. We exploited the complementarity of PALM and SOFI on our focal adhesion data with a methodology integrating a resolution and a signal-to-noise (SNR) metric. Supplementary Note 1 describes the resolution measure, whereas Supplementary Note 3 describes the SNR estimation.

We showed that PALM and SOFI can both be independently applied on the same dataset with additional benefit. Using both methods provides a larger insight into focal adhesions due to their complementarity. To investigate the complementarity of PALM and SOFI under a broader range of controlled conditions, we have introduced a novel approach for assessing contrast and resolution, based on modulation transfer function (MTF) measurements on a simulated test pattern. Supplementary Note 4 describes details about the MTF test pattern and the photophysical model.

Finally, we explored the complementarity of PALM and SOFI for quantitative imaging of focal adhesions. Supplementary Note 5 describes how we corrected blinking events in PALM data in order to obtain the number of emitters and average times between blinking events. Supplementary Note 2 describes how we used SOFI to determine fluorophore densities and activation/switching rates. Supplementary Note 2 additionally describes how we enhanced the SOFI algorithm in order to achieve higher resolution and higher contrast for high order SOFI images. Using enhanced SOFI, we have achieved a spatial resolution comparable to PALM.

## Contents

# Supplementary Note 1: Resolution estimation using sectorial Fourier ring correlation

Estimating the resolution in single molecule localization microscopy (SMLM) is challenging, since it depends on several parameters such as the label density, the localization precision, and the sample structure. An interesting resolution metric for SMLM is the Fourier ring correlation (FRC) [1, 2]. However, the FRC implicitely assumes that the sample structure is isotropic, while focal adhesions are known for their typical pattern giving rise to anisotropic Fourier spectra. We have therefore adapted the FRC metric and introduced sectorial Fourier ring correlation (sFRC) to account for the effect of specific focal adhesion patterns.

## 1.1 From FRC to sFRC

To calculate the FRC, a SMLM dataset is first split into two stochastically independent sub-sets for generating two SMLM images $I_1(x, y)$ and $I_2(x, y)$. Next, the Fourier transforms $\hat{I}_1(q, \phi)$ and $\hat{I}_2(q, \phi)$ of these two images are calculated (with polar coordinates in frequency space given by magnitude $q$ and phase $\phi$). The FRC is then calculated as

$$FRC(q) = \frac{\sum_\phi \hat{I}_1(q, \phi)\hat{I}_2(q, \phi)^*}{\sqrt{\sum_\phi |\hat{I}_1(q, \phi)|^2 \sum_\phi |\hat{I}_2(q, \phi)|^2}} \tag{1}$$

correlating $\hat{I}_1(q, \phi)\hat{I}_2(q, \phi)^*$ over a full circular path at a constant magnitude $q$. For low spatial frequencies, the FRC is close to 1, whereas for high spatial frequencies, the FRC decays to 0. Finally, after applying a smoothing step, the FRC resolution can be calculated as the inverse of the radial frequency for which the curve drops below 1/7 (i.e. the radial cut-off frequency), as suggested in [2].

Evaluating the cross-correlation of $\hat{I}_1(q, \phi)$ and $\hat{I}_2(q, \phi)$ along a circular path entails an insensitivity to pronounced directional variations in the spatial frequency content, as shown in Supplementary Figure 1. This occurs especially for our images containing pronounced specific patterns of cell adhesions, which have a strongly directional imbalanced Fourier spectra due to this adhesion pattern (see Supplementary Figure 1a). We therefore introduced a generalization of the FRC, named sectorial FRC (sFRC). As already suggested in [2], with this generalized metric, the correlation taken over a full circle is replaced by the correlation over a sector with an angular extend of $\Delta\phi$:

$$sFRC(q, \Delta\phi) = \frac{\sum_{\Delta\phi} \hat{I}_1(q, \phi)\hat{I}_2^*(q, \phi)}{\sqrt{\sum_{\Delta\phi} |\hat{I}_1(q, \phi)|^2 \sum_{\Delta\phi} |\hat{I}_2(q, \phi)|^2}} \tag{2}$$

This sFRC concept takes into account major anisotropy of the image spectrum (see Supplementary Figure 1 and 11). Obviously, the classical FRC metric is easily recovered by calculating the sFRC for a sector with an angular extend of $2\pi$. As a guideline, we suggest to evaluate the sFRC in 12 sectors with an angular extend of $\pi/12$, compromising between an improvement in sampling of the direction and a decrease in the amount of data.

## 1.2 Selecting two stochastically independent data subsets

The (s)FRC calculation requires the SMLM data to be split in two stochastically independent subsets, in order to render two stochastically independent SMLM images. First, the full image sequence that consists of $N$ frames is split into $K$ subsequences (containing $N/K$ frames) with $K$ an even number.

Next, $K$ subsequences are split into two subsets of these $K$ subsequences (as shown in Supplementary Figure 8). Each of the two subsets is used to generate one super-resolved image for the (s)FRC calculation. The selection of subsequences into 2 subsets should be done carefully. In case of a random selection, the second subset may contain almost no SMLM data due to photobleaching. This problem can be largely avoided by selection of $K$ subsequences in an alternating way, i.e. creating two subsets of odd and even image subsequences (Supplementary Figure 8).

## 1.3 sFRC applied to SOFI

Although the FRC metric has been conceived for SMLM, we also applied the (s)FRC metric to SOFI data. After splitting the full image sequence into $K$ subsequences, a SOFI image needs to be calculated for each subsequence. After selecting a first set of $K/2$ subsequences, a SOFI image is obtained by summing the SOFI images corresponding to these subsequences. Applying the same procedure to the second $K/2$ subsequences yields two SOFI images. As the SOFI analysis requires consecutive frames, subsequences of limited number of frames should be chosen. We suggest to choose subsequences that contain at least $N/K = 500$ frames.

# Supplementary Note 2: Enhanced bSOFI and on-time ratio estimation

Cell adhesion and its dynamics have been for the first time assessed with SOFI up to the 6th order. Besides a gain in resolution we also addressed the impact on contrast. Using bSOFI up to the sixt order translates into a decrease in contrast necessitating to readdress this question to fully exploit the potential of bSOFI imaging.

We modified the bSOFI algorithm [3] by introducing a linearization for a better compensation of the intrinsic nonlinearity of SOFI. This takes into account in addition the on-time ratio and linearizes the response to detected intensity. In summary, we show that this step improves the attainable resolution assessed by the sFRC as well as the contrast (Supplementary Figure 13).

## 2.1 SOFI theory

SOFI is based on higher order statistics i.e. spatial-temporal cross-cumulants calculated from a time series of images of stochastically blinking emitters. The resolution improvement is given by the properties of these cumulants and described in a seminal paper by J. Enderlein and coworkers[4].

Assuming N independently fluctuating emitters, the detected intensity can be described as

$$I(\mathbf{r},\mathbf{t}) = \sum_{k=1}^{N} \epsilon_k U(\mathbf{r}-\mathbf{r}_k) s_k(t) + b(\mathbf{r}) + n(\mathbf{r},t) \tag{3}$$

where $\epsilon_k$ is the molecular brightness, $U(\mathbf{r}-\mathbf{r}_k)$ is the PSF of the optical system, $s_i(t)$ denotes a switching function (normalized fluctuation sequence, $s_k(t) \in \{0,1\}$), $b(\mathbf{r})$ is a constant background, and $n(\mathbf{r},t)$ represents an additive noise. The sample is assumed to be stationary during the image acquisition. Generally spatial-temporal cumulants can be calculated with various time lags. For reducing the computational complexity and ensuring the maximum of the signal, we used zero time lags. As shown in [5], virtual pixels can be calculated in between the physical pixels acquired by the camera using cross cumulants and followed by a flattening operation (i.e. assigning proper weights to virtual pixels) [5, 6, 7].

Using the properties of cumulants, the n-th order cumulant with zero time lag applied to Eq.3 can be written as

$$\kappa_n\{I(\mathbf{r},\mathbf{t})\} = \sum_{k=1}^{N} \epsilon_k^n U^n(\mathbf{r}-\mathbf{r}_k)\kappa_n\{s_k(t)\} + \kappa_n\{b(\mathbf{r})\} + \kappa_n\{n(\mathbf{r},t)\} \tag{4}$$

For $(n \geq 2)$, under the assumption of uncorrelated noise and stationary background, the terms $\kappa\{b(\mathbf{r})\}$ and $\kappa\{n((r),t)\}$ will cancel out.

In the PALM photophysical model, the emitter activation is taken as non-reversible, however since the emitter is activated, it exhibits several quick blinking events prior to be finally bleached [8]. On a shorter time scale (within one subsequence of input dataset), the emitter fluctuates. If the emitter fluctuates between two different states (bright state $S_{on}$ and dark state $S_{off}$), we can define the on-time ratio as

$$\rho = \frac{\tau_{on}}{\tau_{on} + \tau_{off}} \tag{5}$$

where $\tau_{on}$ and $\tau_{off}$ are the characteristic lifetimes of $S_{on}$ and $S_{off}$ states. The n-th order cumulant $\kappa_n\{s_k(t)\}$ is in this model described by a Bernoulli distribution with probability $\rho_{on}$ [3] and approximated by an n-th order polynomial function of the on-time ratio (further referred to as a cumulant function)

$$f_n(\rho_{on}) = \rho_{on}(1-\rho_{on})\frac{\partial f_{n-1}}{\partial \rho_{on}} \tag{6}$$

Under these conditions, the n-th order spatial-temporal cross-cumulant can be approximated as

$$\kappa_n\{I(\mathbf{r},\mathbf{t})\} \approx \epsilon^n f_n(\rho_{on}) \sum_{k=1}^{N} U^n(\mathbf{r}-\mathbf{r}_k) \tag{7}$$

## 2.2 Linearization and higher order SOFI

The molecular brightness as described in Eq. 7 is raised to the n-th power. High order cumulant images exhibit fluorescent spots of high brightness which are masking less bright structural details. The non-linear response to molecular brightness limits the use of high order cumulants with consequences on resolution enhancement and contrast. S. Geissbuehler et al. proposed balanced SOFI (bSOFI) which allows one to linearize the nonlinear brightness response. Firstly, the n-th order cumulant image is deconvolved. Secondly, the brightness response is linearized by taking the n-th root of the deconvolved cumulant image. This approach has proven efficient for 2D and 3D super-resolution imaging [3, 9].

When using SOFI up to the sixth order, we need to readdress the linearization by taking into account the contribution of $f_n(\rho_{on})$ in Eq. 7. Figure 12 shows the cumulant function dependence on the on-time ratio $\rho_{on}$ for different orders. In the case of a 4th order cumulant and $\rho_{on} = 0.2$, the cumulant function decreases. Under these conditions, the contrast of the 4th order cumulant image is attenuated. The resulting image is flat and the dynamic range is reduced strongly which leads to a loss of SNR. In general, the SNR drops with increasing orders limiting the maximum available resolution enhancement. To overcome this problem, we introduced a novel linearization procedure which takes into account the influence of the cumulant function and linearizes the response to the detected intensity.

The first four cumulants can be written as

$$\begin{aligned}
g_1 &\approx \epsilon(\mathbf{r}) f_1(\rho_{on}) \sum_{k=1}^{N} U(\mathbf{r}-\mathbf{r}_k) + \kappa_1\{b(\mathbf{r})\} + \kappa_1\{n(\mathbf{r},t)\} \\
g_2 &\approx \epsilon^2(\mathbf{r}) f_2(\rho_{on}) \sum_{k=1}^{N} U^2(\mathbf{r}-\mathbf{r}_k) \\
g_3 &\approx \epsilon^3(\mathbf{r}) f_3(\rho_{on}) \sum_{k=1}^{N} U^3(\mathbf{r}-\mathbf{r}_k) \\
g_4 &\approx \epsilon^4(\mathbf{r}) f_4(\rho_{on}) \sum_{k=1}^{N} U^3(\mathbf{r}-\mathbf{r}_k)
\end{aligned} \tag{8}$$

and the on-time ratio polynomials up to the sixth order are

$$f_1(\rho_{on}) = \rho_{on} \tag{9}$$
$$f_2(\rho_{on}) = \rho_{on}(1-\rho_{on}) \tag{10}$$
$$f_3(\rho_{on}) = \rho_{on}(1-\rho_{on})(1-2\rho_{on}) \tag{11}$$
$$f_4(\rho_{on}) = \rho_{on}(1-\rho_{on})(1-6\rho_{on}+6\rho_{on}^2) \tag{12}$$
$$f_5(\rho_{on}) = \rho_{on}(1-\rho_{on})(1-2\rho_{on})(12\rho_{on}^2-12\rho_{on}+1) \tag{13}$$
$$f_6(\rho_{on}) = \rho_{on}(1-\rho_{on})(120\rho_{on}^4-240\rho_{on}^3+150\rho_{on}^2-30\rho_{on}+1) \tag{14}$$

Once the on-time ratio is estimated (as described in the next section), the value of the on-time ratio polynomial for a given cumulant order is calculated by Eq. 10-14. In order to correct for the amplified brightness without compromising the resolution, the cumulants have to be deconvolved first as shown in [3]. The correction factor for a deconvolved n-th order cumulant image $\hat{g}_n$ is $1/f_n(\rho_{on})$ and we can write

$$\frac{\hat{g}_n}{f_n(\rho_{on})} = \hat{g}_n^{\frac{log_{10}(\hat{g}_n/f_n(\rho_{on}))}{log_{10}(\hat{g}_n)}} \tag{15}$$

Instead of taking the n-th root, the corrected, adaptively linearized cumulant image $\bar{g}_n$ is

$$\bar{g}_n = g_n^{\frac{1}{n}\frac{log_{10}(g_n/f_n(\rho_{on}))}{log_{10}(g_n)}} \tag{16}$$

The roots for linearization of cumulants up to 6th order (linearization curve) and the difference in the final bSOFI images are shown in Figure 13. The red line in Figure 13a represents the standard linearization where the n-th order cumulant is linearized by taking the n-th root [10]. The corrected roots for our novel linearization are shown in blue (Figure 13a).

## 2.3 On-time ratio estimation

Higher-order cumulants contain information about the photo-physics of the emitters. Combining SOFI images of different cumulant orders, molecular parameter maps can be extracted such as on-time ratio, molecular brightness, and molecular density [3], which we applied to assess the dynamics of cell adhesions. Geissbuehler et al. [3] used three cumulant images (2$^{\text{nd}}$, 3$^{\text{rd}}$, and 4$^{\text{th}}$ order) to estimate the on-time ratio. Here we present an estimation of the on-time ratio using only two cumulant images (2$^{\text{nd}}$ and 3$^{\text{rd}}$ order). If we assume spatially varying but locally constant on-time ratios and molecular brightness, the cumulants can be approximated by [3]

$$
\begin{aligned}
g_1(\mathbf{r}) &\approx \epsilon(\mathbf{r}) f_1(\rho_{on}) N(\mathbf{r}) \mathcal{E}_V\{U(\mathbf{r})\} + \kappa_1\{b(\mathbf{r})\} + \kappa_1\{n(\mathbf{r},t)\} && (17)\\
g_2(\mathbf{r}) &\approx \epsilon^2(\mathbf{r}) f_2(\rho_{on}) N(\mathbf{r}) \mathcal{E}_V\{U^2(\mathbf{r})\} && (18)\\
g_3(\mathbf{r}) &\approx \epsilon^3(\mathbf{r}) f_3(\rho_{on}) N(\mathbf{r}) \mathcal{E}_V\{U^3(\mathbf{r})\} && (19)\\
g_4(\mathbf{r}) &\approx \epsilon^4(\mathbf{r}) f_4(\rho_{on}) N(\mathbf{r}) \mathcal{E}_V\{U^4(\mathbf{r})\} && (20)\\
& && (21)
\end{aligned}
$$

where $\mathcal{E}_V\{U^n(\mathbf{r})\}$ is the expectation value of $U^n(\mathbf{r})$, $N(\mathbf{r})$ is the number of molecules inside a detection volume V centered at $\mathbf{r}$. The second ($g_2$) and third ($g_3$) order cumulant images can be related as

$$
g_3 = \frac{\mathcal{E}_V\{U^3(\mathbf{r})\}}{\mathcal{E}_V\{U^2(\mathbf{r})\}^{3/2}} \frac{1}{N^{1/2}(\mathbf{r})} \frac{f_3(\rho_{on})}{f_2^{3/2}(\rho_{on})} g_2^{3/2} \tag{22}
$$

Substituting Eq.10 and Eq. 11 into Eq. 22 leads to

$$
g_3 = K \frac{1 - 2\rho_{on}}{\sqrt{\rho_{on}(1-\rho_{on})}} g_2^{3/2} \tag{23}
$$

where $K = \frac{\mathcal{E}_V\{U^3(\mathbf{r})\}}{\mathcal{E}_V\{U^2(\mathbf{r})\}^{3/2}} \frac{1}{N^{1/2}(\mathbf{r})}$.
For the on-time ratio $\rho_{on}$, we obtain the solutions

$$
\rho_{on} = \frac{1}{2}\left(1 \pm \frac{\sqrt{4K^2 g_2^3 g_3^2 + g_3^4}}{4K^2 g_2^3 + g_3^2}\right) \tag{24}
$$

As shown in Figure 12, the on-time ratio polynomial is symmetric around $\rho_{on} = 0.5$, thus the Eq. 24 has two possible solutions. To estimate $\rho_{on}$, we first determine the constant $K$. The number of molecules $N(\mathbf{r})$ can be estimated using the second order cumulant and the first order cumulant after background subtraction ($\tilde{g}_1$).

$$
N(\mathbf{r}) = \frac{\mathcal{E}_V\{U^2(\mathbf{r})\}}{\mathcal{E}_V\{U^1(\mathbf{r})\}^2} \frac{(1-\rho_{on})}{\rho_{on}} \frac{\tilde{g}_1}{g_2} \tag{25}
$$

Approximating the imaging PSF by a 3D Gaussian profile, we can write [3]

$$
\mathcal{E}_V\{U^n_{\text{3DGauss}}(\mathbf{r})\} = \frac{c(\sigma_{x,y}, \sigma_z)}{n^{3/2}} \tag{26}
$$

where $c(\sigma_{x,y}, \sigma_z)$ is a constant depending on the spatial extend of the PSF. Analogously, approximating the PSF near the interface in a total internal reflection (TIR) configuration by a lateral 2D Gaussian profile and an axial exponential profile, we obtain

$$
\mathcal{E}_V\{U^n_{\text{TIR}}(\mathbf{r})\} = \frac{c(\sigma_{x,y}, \sigma_z, d_z)}{n^2} \tag{27}
$$

where $d_z$ represents the penetration depth of the TIR illumination [3]. The outcome of this analysis has been implemented into our SOFI code inducing the expected contrast gain.

To show the accuracy of the above described on-time ratio estimation, we performed testing on simulated data. Figure 14 shows the results with varying number of frames and 2, 5, 10, and 20 molecules randomly distributed within the PSF volume. The simulation for each reference on-time ratio was repeated 20 times. Each estimated on-time ratio is an average over these 20 calculations.

Density and molecular on-time ratio maps in Figure 4 were calculated by taking the Eq. 17, 18, 19 and solving this system of equations pixel-wise for $\epsilon(\mathbf{r})$, $\rho_{on}$, and $N(\mathbf{r})$ as described in [10]. Figure 4 shows color-coded density and molecular on-time ratio overlaid with the 4th order bSOFI image as a transparency mask. This pixel-wise estimation is not relevant for image regions which contain only background noise. Therefore the linearized SOFI image is used as a transparency mask to cancel out the background regions. The bSOFI image was linearized using our novel adaptive linearization procedure described in the section 5.3.

# Supplementary Note 3: Signal-to-noise ratio estimation using statistical resampling

Imaging dynamics of cell adhesions trades spatial against temporal resolution with an impact on SNR. Therefore, we characterized the SNR in order to ensure a sufficient image quality. We estimated the SNR in our focal adhesion images by making use of a general approach based on statistical resampling applied to SOFI as well as PALM.

## 3.1 Delete-1 jackknife resampling

The SNR of SOFI images can be calculated by delete-1 jackknife resampling [11], i.e. $N$ new datasets equal to the number of raw images $N$ of the original dataset are generated, but each new dataset leaves out a single image in these sequences, as shown in Supplementary Figure 1b. Each new dataset is used to generate a new SOFI image, yielding $N$ new SOFI images. For each pixel value $I(x, y)$ of the original SOFI, $N$ new $I_n(x, y)$ values are generated. The level of uncertainity associated to each pixel $I(x, y)$ can be quantified using the SNR per pixel, defined as

$$\mathrm{SNR(x,y)} = \frac{I(x,y)}{\sqrt{\mathrm{var}\{I(x,y)\}}} \tag{28}$$

The jackknife mean estimator is

$$\bar{I}(x,y) =< I_n(x,y) > . \tag{29}$$

The jackknife variance estimator is

$$\mathrm{var}\{I(x,y)\} = (N-1) < (I_n(x,y) - \bar{I}(x,y))^2 > . \tag{30}$$

## 3.2 SNR estimation on SOFI data

When calculating SOFI for long image sequences, photobleaching cannot be neglected. The full image sequence is therefore divided into short subsequences during which the photobleaching effect is insignificant. In our case each subsequence contained 500 frames. For decreasing the computational burden while evaluating the jackknife resampling, the SOFI image is first pre-calculated for each subsequence. The resampling is always performed within one subsequence, then the pre-calculated SOFI images from the remaining subsequences are added to generate a new resampled SOFI image, as shown in Supplementary Figure 9. At the beginning, the algorithm takes the first subsequence (the first 500 frames) from a total number of $K$ subsequences. The first frame from this subsequence is discarded. A SOFI image $s_1$ is calculated from the rest of the subsequence (i.e. the following 499 frames). The SOFI image $s_1$ is summed with the pre-calculated SOFI images from the remaining $K-1$ subsequences which yields a resampled SOFI image $I_1(x, y)$. In the next step, the second frame is discarded, leaving a different subset of 499 frames used to calculate a SOFI image $s_2$. Combining $s_2$ with the pre-calculated SOFI images from the remaining $K-1$ subsequences yields a resampled SOFI image $I_2(x, y)$ . When the whole first subsequence is resampled, the procedure is repeated step by step for every subsequence to cover the full image sequence (i.e. 20,000 frames in our data).

## 3.3 SNR estimation on PALM data

Although originally introduced for SOFI, the SNR can also be determined for SMLM data, since SMLM images can be rendered in a pixelated fashion (e.g. as a 2D histogram). Moreover, the SMLM localization procedure does not need to be repeated $N$ times. It would even seem sufficient to localize the molecules only once from the original dataset, and afterwards just rendering $N$ SMLM images by removing the localizations that correspond to the frame that is "deleted". However, there is one caveat: the same emitter can appear during several consecutive frames. This means that deleting its localization when one of these frames is deleted, is not necessarily correct if one imposes an upper limit on the localization precision. The reason is that the localization precision could still be sufficiently small for the localization to be included, based on the contributions from the other frames that were not deleted. Conversely, new localizations can arise by deleting a frame if an upper limit on the localization precision is imposed (e.g. to exclude bright fiduciary markers). In this case, there is a chance that the localization precision becomes sufficiently large upon deleting one of the frames where it was visible. Both problems can be solved by re-estimating the localization precision after the deletion of one frame, as shown in Supplementary Figure 8b. This can be done by making two approximations: (1) the number of photons in each frame is constant, and (2) the localization precision is inversely related to the square root of the amount of photons. An emitter with localization precision $\sigma$ that appeared in $n$ frames therefore obtains a new localization precision after deleting one frame given by

$$\sigma_{\text{delete-1}} = \sigma\sqrt{n-1} \tag{31}$$

After re-calculating the localization precisions and applying the upper and lower limit on the localization precision, the $N$ new SMLM images for the SNR calculation can be rendered.

To calculate the variance in Eq. 28, a sufficient number of localizations have to be present inside the pixel area. If not the case, for instance due to a too small pixel size or a too low localization density, the SNR value can become unreliable.

### 3.4 SNR convergence rate

The SNR increases with an increasing number of frames used for the super-resolution image reconstruction. Supplementary Figure 6 shows the SNR ramp up, i.e. how quickly the SNR converges to a value that remains almost stable at a plateau with almost no further improvement due to an increasing number of frames. The relative increment of the SNR shown in Supplementary Figure 6 was calculated as

$$\text{ISNR} = \frac{|\text{SNR}_{n+1} - \text{SNR}_n|}{|SNR_n|} \tag{32}$$

The measured SNR values are shown in Figure 1.

# Supplementary Note 4: Resolution based on MTF analysis

Modulation transfer function (MTF) analysis on simulated data allows us to compare the spatial-temporal resolution of PALM and SOFI under controlled conditions close to the conditions in focal adhesions.

Comparing PALM and SOFI is challenging due to their very different nature (i.e. a list of localizations vs. higher order statistics calculated across the input image stack). Measures like precision, recall or accuracy are often used when comparing PALM algorithms. In this case, a list of localized emitters is compared with the ground truth data. This approach is not well suited for comparing PALM and SOFI. Although the image resolution improves with increasing SOFI order, SOFI does not provide the localizations of underlying emitters.

Therefore, we propose a new approach based on the MTF analysis using a simulated test pattern. We extended the MTF analysis, already well known from classical optics, for application in super-resolution imaging. This generalized MTF analysis uses the same terminology like the visibility and the cut-off frequency for super-resolution and allows one to assess the full path from object to super-resolved image.

## 4.1 Test pattern

The test pattern is composed of bars with varying width ranging from 500 nm to 20 nm. More precisely, the bars were 500, 400, 300, 200, 150, 120, 100, 90, 80, 70, 60, 50, 40, 30, and 20 nm wide. Repeating every width for three consecutive bars led to the test pattern with 45 bars as shown in Figure 2a. Assuming fluorescent labelling, the bars are filled by uniformly distributed emitters according to a predefined labeling density.

## 4.2 Simulations and photophysical model

The simulation assumes photokinetics typical for fluorescent proteins in PALM measurements. For each fluorophore, a time trace is modelled. The time trace describes number of photons emitted by a given fluorophore over time. Each fluorophore, once it is in the on-state, shows a "burst" of blinking events before being bleached. The blinking fluorophore randomly switches between the on-state and a dark state. On- and off-times of these blinking events, as well as bleaching of the fluorophore, are governed by an exponential distribution with an average on-time $\tau_{\mathrm{on}}$, an average off-time $\tau_{\mathrm{off}}$, and an average bleaching lifetime $\tau_{\mathrm{bl}}$. The on-time ratio $(\tau_{\mathrm{on}}/(\tau_{\mathrm{on}} + \tau_{\mathrm{off}})$ defines the frequency of the fast blinking in the burst. Assuming a camera frame rate of 50Hz, the blinking parameters were set in order to obtain a similar behaviour as mEos2 measured in [8]. The average duration of 8 blinking events in one burst takes in average 3.2 s (on-time ratio = 0.05). The exposure time is assumed to be faster than $\tau_{\mathrm{on}}$ and $\tau_{\mathrm{off}}$ and the blinking is therefore properly sampled. Supplementary Figure 10a shows time traces of the first 10 fluorophores. Please note that the Supplementary Figure 10a shows the time traces before adding noise. The number of blinks per burst is random (in the range 2-10). Supplementary Figure 10c,d shows the statistics of the simulated image stack. The average number of blinks per burst equals to 5.9. Supplementary Figure 10c shows the number of photons as a function of frame number normalized to one. Exponential decay was fitted to measure the average bleaching lifetime. Bleaching in the simulation was set to match our experimentally measured data.

## 4.3 Modulation transfer function (MTF) and sFRC

For our MTF analysis, the pattern, consists of progressively narrower black and white bars. When imaging this pattern the bars might still be resolved, but the visibility decreases with increasing spatial frequencies. The visibility is given as

$$M = (F_{\mathrm{max}} - F_{\mathrm{min}})/(F_{\mathrm{max}} + F_{\mathrm{min}}) \tag{33}$$

where $F_{\mathrm{max}}$ and $F_{\mathrm{min}}$ are taken as the maximum and minimum intensity values at a given spatial frequency. In the classical optics, the microscope is described as a low pass filter. The MTF describes this filtering effect when comparing a periodic object (with a givne spatial frequency) to the filtered image. The MTF can easily be calculated as

$$\mathrm{MTF} = |\mathcal{F}\{P(\mathbf{r})\}| \tag{34}$$

where $P(\mathbf{r})$ represents the test pattern and the operator $\mathcal{F}\{\}$ corresponds to the Fourier transform. The modulation depth is associated to the afformentioned visibility and this generalized MTF analysis integrates all contributions starting from the object and ending with the super-resolved image. Assuming no noise, the cut-off frequency $f_c$ corresponds to the spatial frequency where the visibility goes to zero and the limit of resolution is given by $1/f_c$.

The resolution measured by the MTF analysis (rMTF) can be related to the resolution based on sFRC. The test pattern described in section 4.1 is a strongly rectangular object with spatial frequencies changing along one direction. These conditions are not suitable for the standard FRC calculation. Therefore, we have calculated the sFRC on a segment in the direction which corresponds to the main spectral content in the Fourier space. Supplementary Figure 4ab shows a comparison of the sFRC and rMTF for labeling density 800 and 1200 molecules/$\mu m^2$, respectively. Interpretation of the sFRC values should be done carefully, as the spatial resolution value obtained from the MTF analysis is typically slightly lower.

# Supplementary Note 5: Correcting blinking events in PALM data

PALM can be used to obtain quantitative molecular information of focal adhesions, such as the number of fluorescent proteins and the blinking off times. However, simply counting the localizations usually yields an overestimation of this quantity, since fluorescent proteins are known to blink. This error can be avoided by merging localizations that are close in time and space. However, applying such counting methods on focal adhesion data is challenging, since focal adhesions are dense protein structures. We have, therefore, adapted a counting method to take higher localization densities into account.

## 5.1 Spatial and temporal clustering of blinking events

Counting blinking fluorescent proteins from PALM data can be done as published in [12]. If two different localizations $x_a$ and $x_b$ are sufficiently close and observed within a sufficiently small time interval, they can be assumed to originate from two blinking events of the same fluorescent protein. First, all localization pairs with a time interval below a certain threshold value $t_{\mathrm{d}}$ are considered as potential blinking events. Next, a second selection is made based on the distance between them, i.e. they are considered blinking events if they are closer than a distance threshold. After merging, the localizations are again evaluated against the same criteria until no blinking events can be identified. In order to apply the method to correct for blinking, the value of $t_{\mathrm{d}}$ is varied in multiples of the camera exposure time $t_{\mathrm{exp}}$. For each $t_{\mathrm{d}}$ the total number of localizations $N(t_{\mathrm{d}})$ is determined, and these values are fitted to the following semi-empirical model in order to obtain the correct number $N$ of fluorescent proteins [12]

$$N(t_{\mathrm{d}}) = N(1 + n_{\mathrm{blink}} e^{\frac{t_{\mathrm{exp}} - t_{\mathrm{d}}}{t_{\mathrm{off}}}}) \tag{35}$$

where $n_{\mathrm{blink}}$ is the average number of times a fluorescent protein blinks and $t_{\mathrm{off}}$ is the average time between two blinking events. This model assumes that the fluorescent protein first goes from an inactivated to an activated state. Once the protein is activated, it either reversibly goes to a dark state or irreversibly to a photobleached state (see Supplementary Figure 10e). For large values of $t_{\mathrm{d}}$, the model predicts that the observed $N(t_{\mathrm{d}})$ approaches $N$, as would be expected. However, the larger the value of $t_{\mathrm{d}}$, the higher the probability of grouping localizations from different fluorescent proteins, which is not accounted for by the model. Hence, the fit is only performed for small values of $t_{\mathrm{d}}$, i.e. the first 5 multiples of $t_{\mathrm{exp}}$, as suggested in [12].

## 5.2 Distance threshold accounting for localization precision

A single distance threshold value for all localizations should be avoided, since the localization precision $\sigma_a$ and $\sigma_b$ corresponding to $x_a$ and $x_b$, respectively, can be very different. Consider the observed localizations $x_a$ and $x_b$ to be normally distributed around the real protein positions $\mu_a$ and $\mu_b$, respectively, with standard deviation $\sigma_a$ and $\sigma_b$, respectively. The question whether both localizations are originating from the same emitter thus boils down to the question how similar both normal distributions are. We therefore defined a threshold based on the Hellinger distance, a statistical measure that probes the similarity between two normal distributions. The Hellinger distance $H$ can be calculated from the following expression

$$H^2 = 1 - \sqrt{\frac{2\sigma_a\sigma_b}{\sigma_a^2 + \sigma_b^2}} exp(-\frac{1}{4}\frac{(\mu_a - \mu_b)^2}{\sigma_a^2 + \sigma_b^2}) \tag{36}$$

The Hellinger distance varies between 0 and 1. It is equal to 0 if both probability distributions are identical, and it is equal to 1 if the two probability distributions do not overlap. A threshold value of 0.9 is a reasonable choice, since it corresponds to a distance threshold between two localizations equal to $\sim 3.6$ times the localization precision, assuming that their localization precisions are equal. As the real positions are not known, we approximate $\mu_a - \mu_b$ by $x_a - x_b$ in order to calculate the Hellinger distance.

## 5.3 Position and localization precision of merged blinking events

The merging procedure is repeated until no blinks can be identified, so one needs to calculate the position and localization precision of merged blinking events. Assuming that the observed localizations are normally distributed around their real positions with the localization precision as the standard deviation, we consider $\sigma_a$ and $\sigma_b$ as weights to calculate the new position as follows

$$x_{\text{merged}} = \frac{x_a/\sigma_a^2 + x_b/\sigma_b^2}{1/\sigma_a^2 + 1/\sigma_b^2} \tag{37}$$

The corresponding localization precision of the merged position is given by

$$\sigma_{\text{merged}} = \frac{1}{\sqrt{1/\sigma_a^2 + 1/\sigma_b^2}} \tag{38}$$

Although the merging procedure was described in one dimension, its application for two dimensional data was done for each dimension separately, i.e. localizations were considered to be blinking events when they were identified as such in both dimensions.

# References

[1] N. Banterle, K. H. Bui, E. A. Lemke, and M. Beck, "Fourier ring correlation as a resolution criterion for super-resolution microscopy," *Journal of Structural Biology*, vol. 183, pp. 363–367, 2013.

[2] R. P. J. Nieuwenhuizen, K. a. Lidke, M. Bates, D. L. Puig, D. Grünwald, S. Stallinga, and B. Rieger, "Measuring image resolution in optical nanoscopy.," *Nature methods*, vol. 10, pp. 557–62, 2013.

[3] S. Geissbuehler, N. L. Bocchio, C. Dellagiacoma, C. Berclaz, M. Leutenegger, and T. Lasser, "Mapping molecular statistics with balanced super-resolution optical fluctuation imaging (bSOFI)," *Optical Nanoscopy*, vol. 1, no. 1, p. 4, 2012.

[4] T. Dertinger, R. Colyer, G. Iyer, S. Weiss, and J. Enderlein, "Fast, background-free, 3D super-resolution optical fluctuation imaging (SOFI).," *Proceedings of the National Academy of Sciences of the United States of America*, vol. 106, no. 52, pp. 22287–22292, 2009.

[5] T. Dertinger, R. Colyer, R. Vogel, J. Enderlein, and S. Weiss, "Achieving increased resolution and more pixels with Superresolution Optical Fluctuation Imaging (SOFI).," *Optics express*, vol. 18, pp. 18875–85, Aug. 2010.

[6] S. C. Stein, A. Huss, D. Hähnel, I. Gregor, and J. Enderlein, "Fourier interpolation stochastic optical fluctuation imaging.," *Optics express*, vol. 23, pp. 16154–63, 2015.

[7] W. Vandenberg, M. Leutenegger, T. Lasser, J. Hofkens, and P. Dedecker, "Diffraction-unlimited imaging: from pretty pictures to hard numbers," *Cell and Tissue Research*, 2015.

[8] N. Durisic, L. Laparra-Cuervo, A. Sandoval-Álvarez, J. S. Borbely, and M. Lakadamyali, "Single-molecule evaluation of fluorescent protein photoactivation efficiency using an in vivo nanotemplate.," *Nature methods*, vol. 11, pp. 156–62, 2014.

[9] S. Geissbuehler, A. Sharipov, A. Godinat, N. L. Bocchio, P. a. Sandoz, A. Huss, N. a. Jensen, S. Jakobs, J. Enderlein, F. Gisou van der Goot, E. a. Dubikovskaya, T. Lasser, and M. Leutenegger, "Live-cell multiplane three-dimensional super-resolution optical fluctuation imaging," *Nature Communications*, vol. 5, p. 5830, Dec. 2014.

[10] M. Geissbuehler and T. Lasser, "How to display data by color schemes compatible with red-green color perception deficiencies.," *Optics express*, vol. 21, no. 8, pp. 9862–74, 2013.

[11] W. Vandenberg, S. Duwé, M. Leutenegger, B. Krajnik, T. Lasser, and P. Dedecker, "Model-free uncertainty estimation in Stochastical Optical Fluctuation Imaging ( SOFI ) leads to a doubled temporal resolution," vol. 2402, pp. 1347–1355, 2015.

[12] P. Annibale, S. Vanni, M. Scarselli, U. Rothlisberger, and A. Radenovic, "Quantitative Photo Activated Localization Microscopy: Unraveling the effects of photoblinking," *PLoS ONE*, vol. 6, 2011.

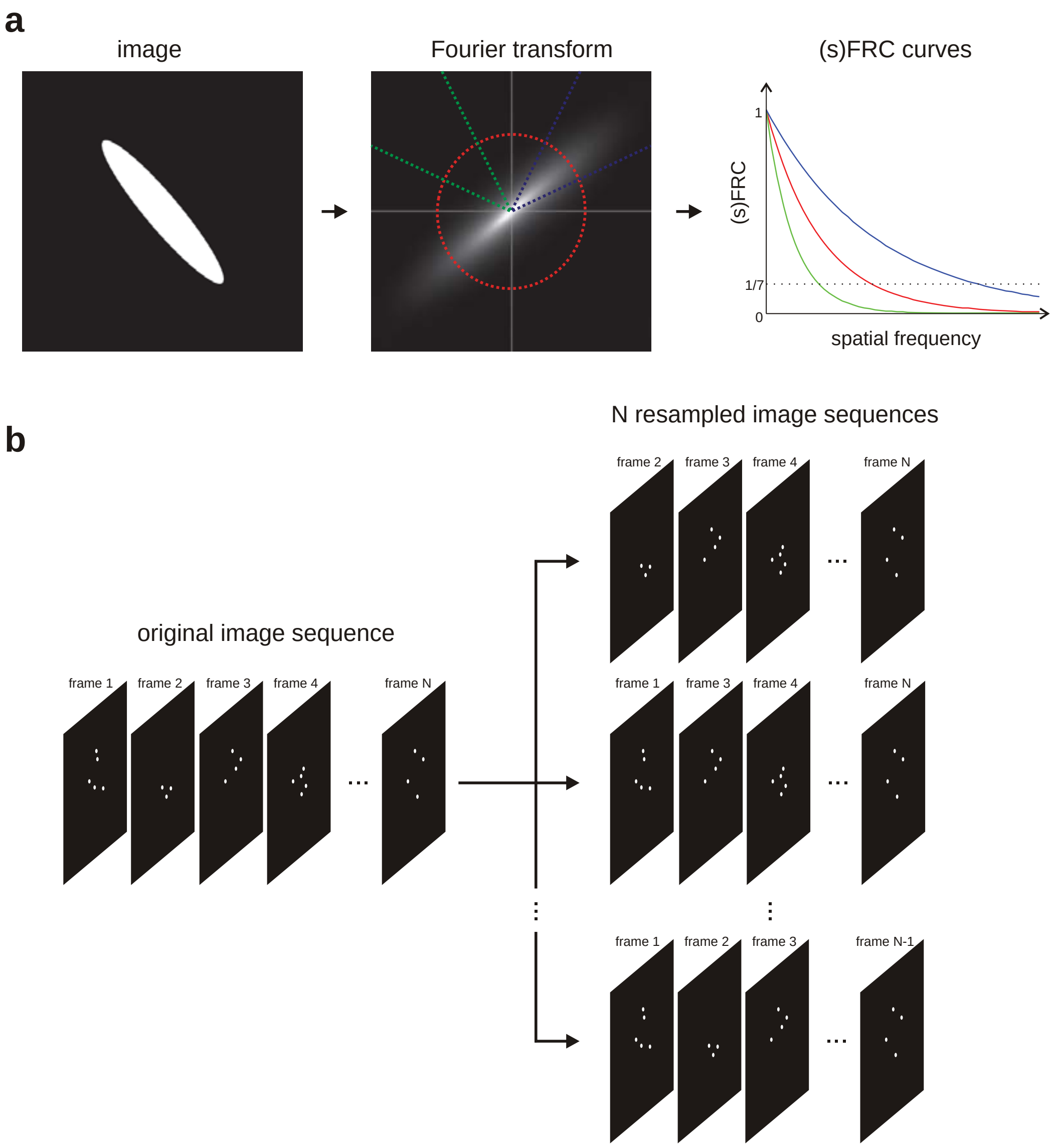

**Supplementary Figure 1:** Resolution and SNR metrics. (a) Illustration of the sectorial Fourier ring correlation to obtain a measure of the resolution. (b) Illustration of the delete-1 jackknife resampling method to obtain a measure for the signal-to-noise ratio.

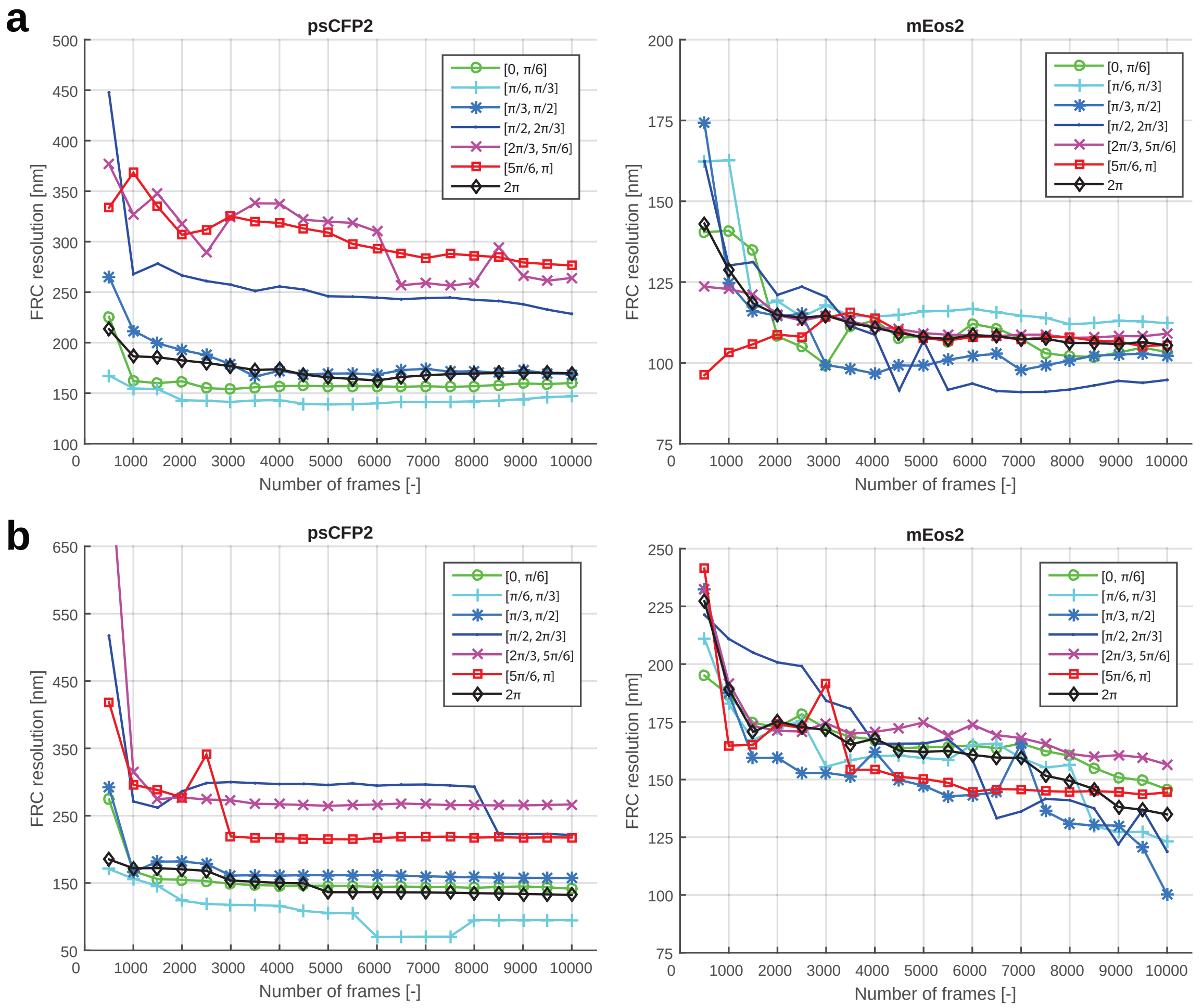

**Supplementary Figure 2:** The sFRC as a function of the number of frames for the data shown in Figure 1. (a) The sFRC for 6 different sectors, together with the FRC, obtained from PALM. (b) The sFRC for 6 different sectors, together with the FRC, obtained from SOFI.

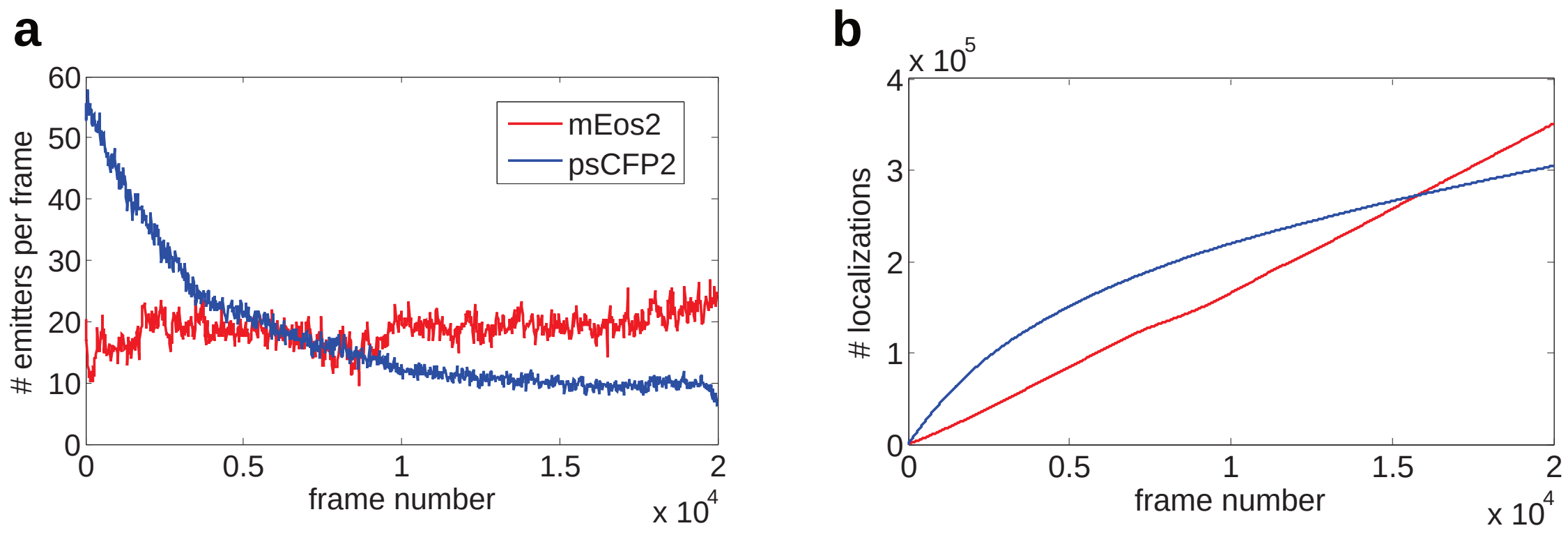

**Supplementary Figure 3:** Time evolution of the number of fluorophores in the data shown in Figure 1. (a) The number of detected emitters per frame (averaged over 20 frames) as a function of the frame number. (b) The number of localizations as a function of the number of frames. Note that an emitter can be detected in several consecutive frames, giving rise to a single localization.

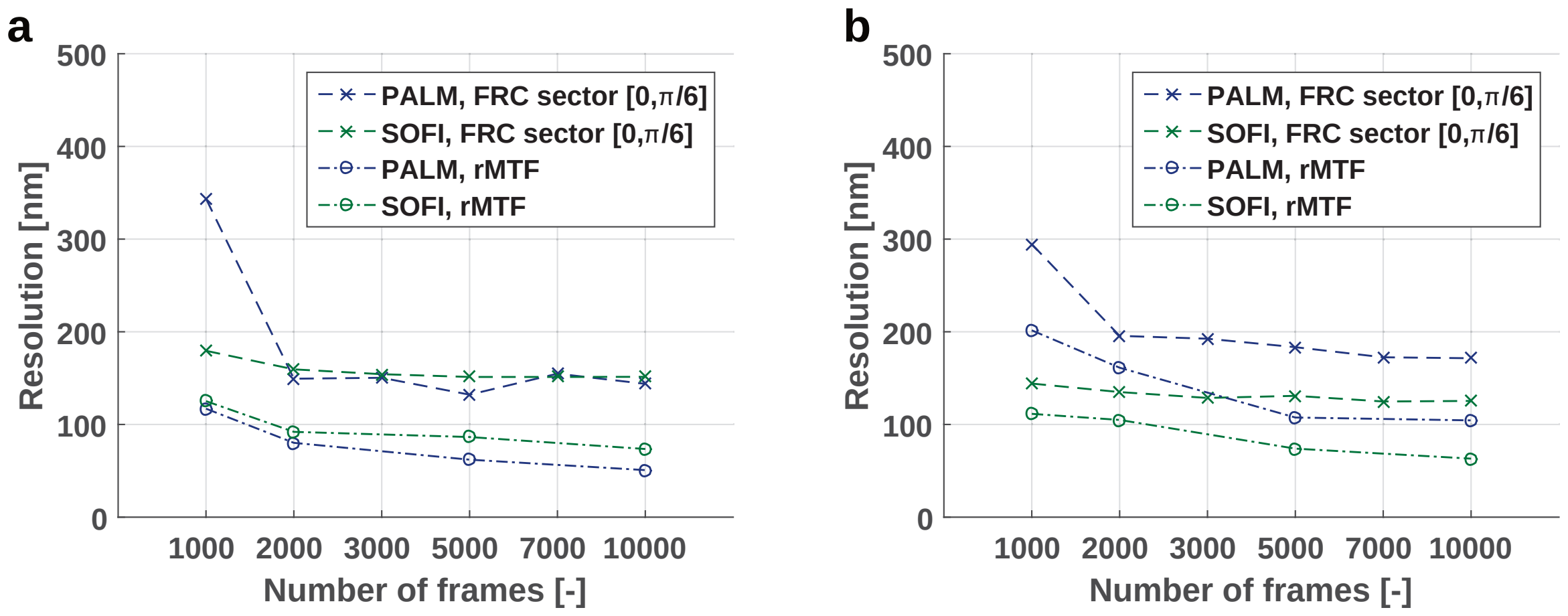

**Supplementary Figure 4:** sFRC and resolution based on MTF both measured on simulated dataset. From all the simulated data sets shown in Figure 2, this Figure corresponds to the dataset with Ion = 100 photons, and 20000 frames. (a) density 800 molecules/$\mu m^2$, (b) density 1200 molecules/$\mu m^2$.

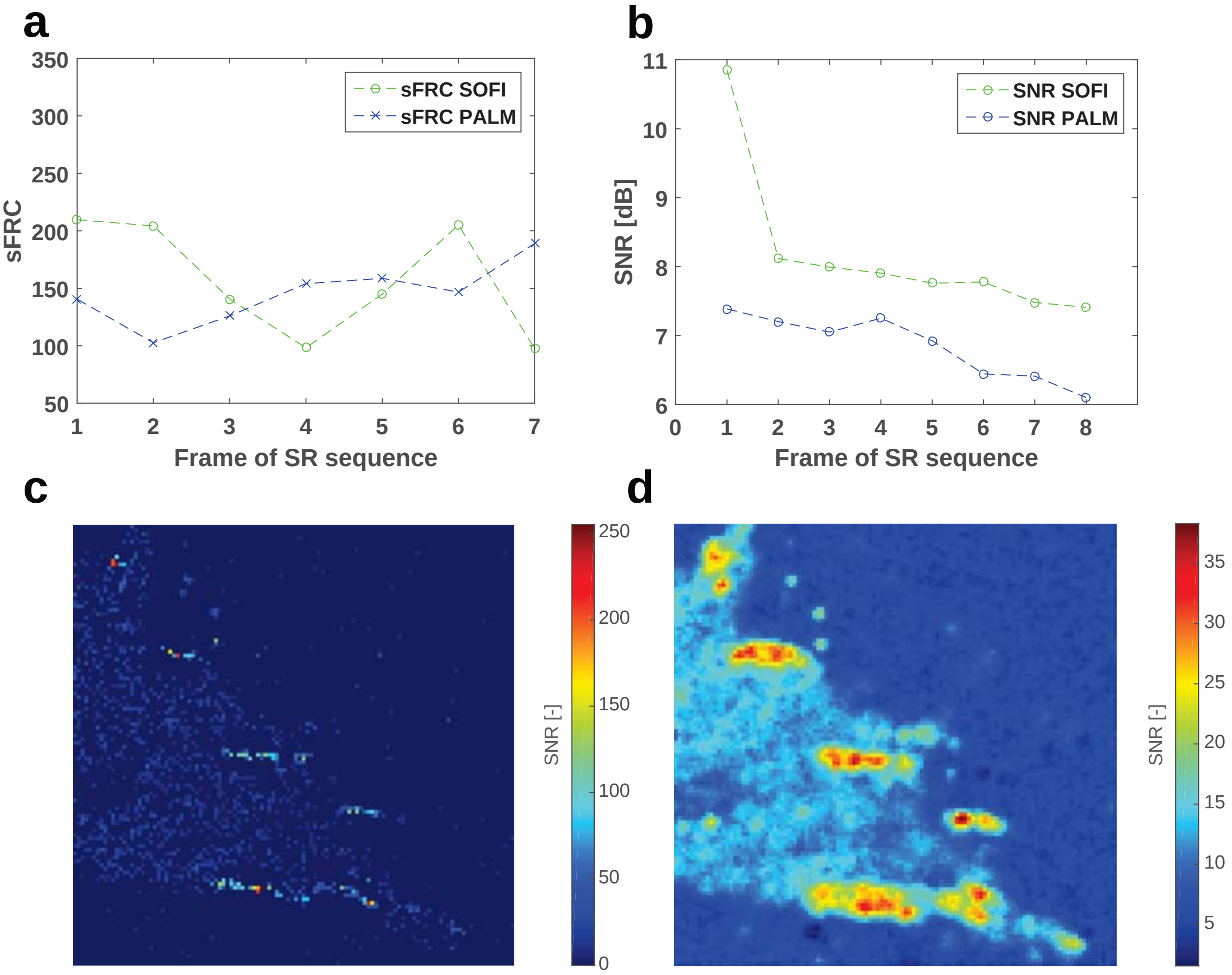

**Supplementary Figure 5:** SNR and sFRC calculated on SOFI and PALM movie of a living MEF expressing paxillin labeled with mEos2. Each image is reconstructed from 1000 camera frames with 10 ms exposure time, resulting in a 10 s temporal resolution. (a) sFRC values for each super-resolved SOFI/PALM frame. (b) average SNR for each super-resolved SOFI/PALM frame. (c) PALM (d) SOFI SNR map of the first frame of the PALM/SOFI output sequence.

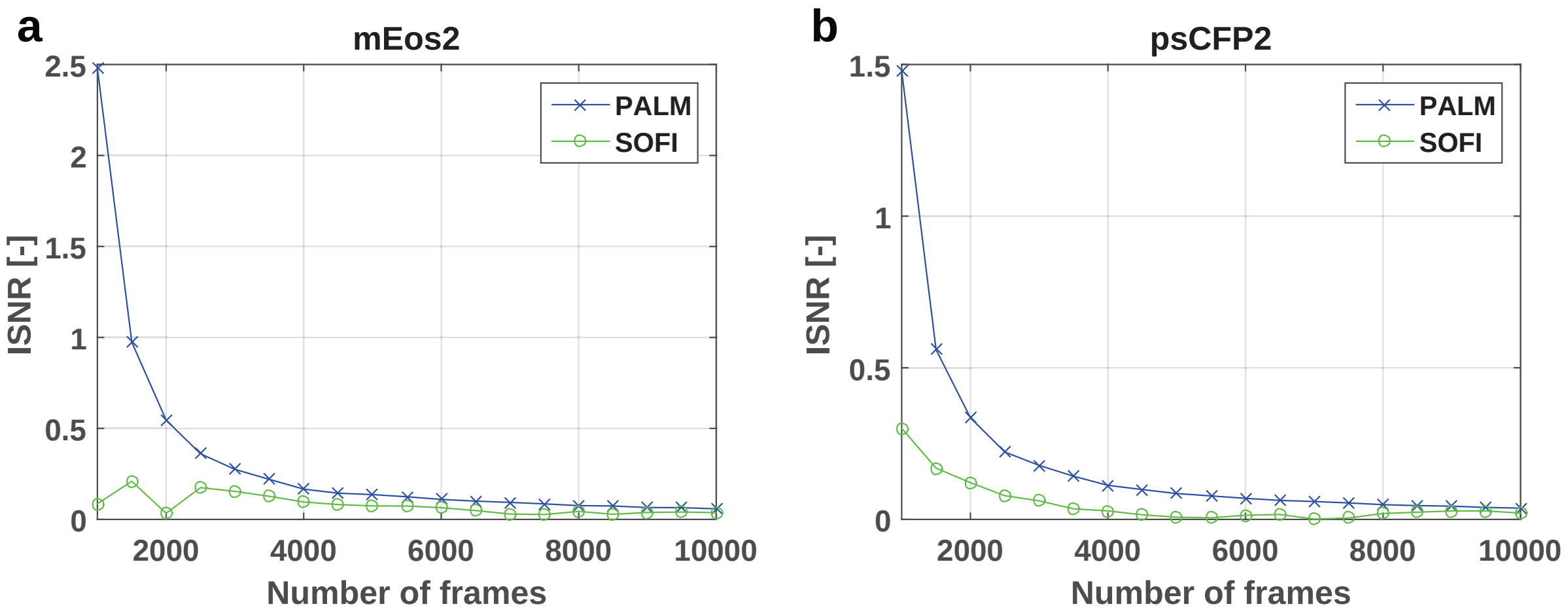

**Supplementary Figure 6:** SNR convergence rate measured on PALM and SOFI images of a fixed MEF expressing paxillin labeled with (a) mEos2 and (b) psCFP2.

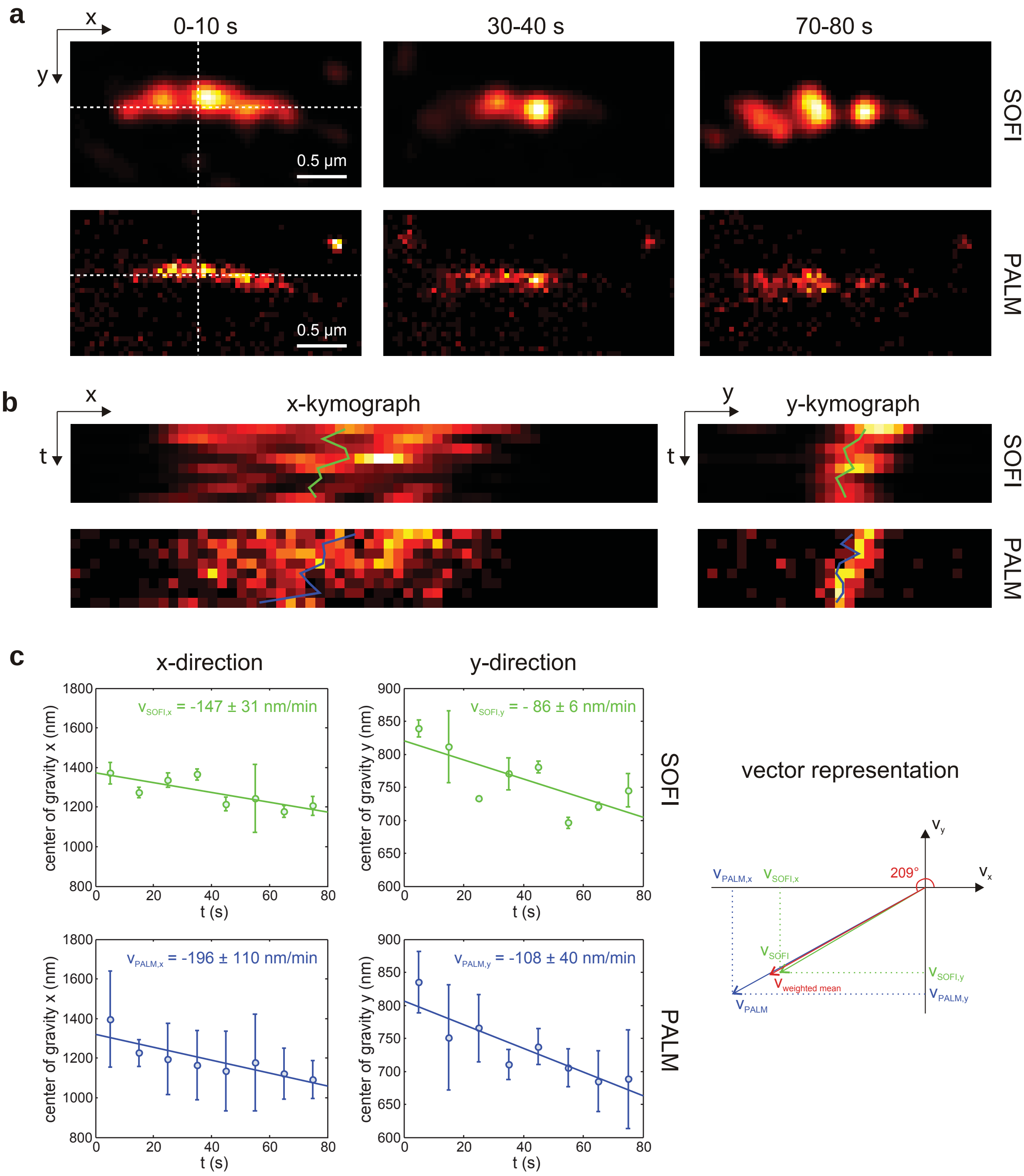

**Supplementary Figure 7:** Live cell imaging with PALM and bSOFI. (a) PALM and SOFI image of living MEF expressing paxillin labeled with mEos2. (b) Region of interest indicated in (a) at different time points. Each image is reconstructed from 1000 camera frames with 10 ms exposure time, resulting in a 10 s temporal resolution. (c) Kymographs along the lines indicated in (b). (d) Focal adhesion edge position as a function of time determined from the kymographs in (c). The edges were identified as the first pixel from both sides in the kymograph with a value that exceeds half of the maximum value. The velocities have been obtained by a linear fit.

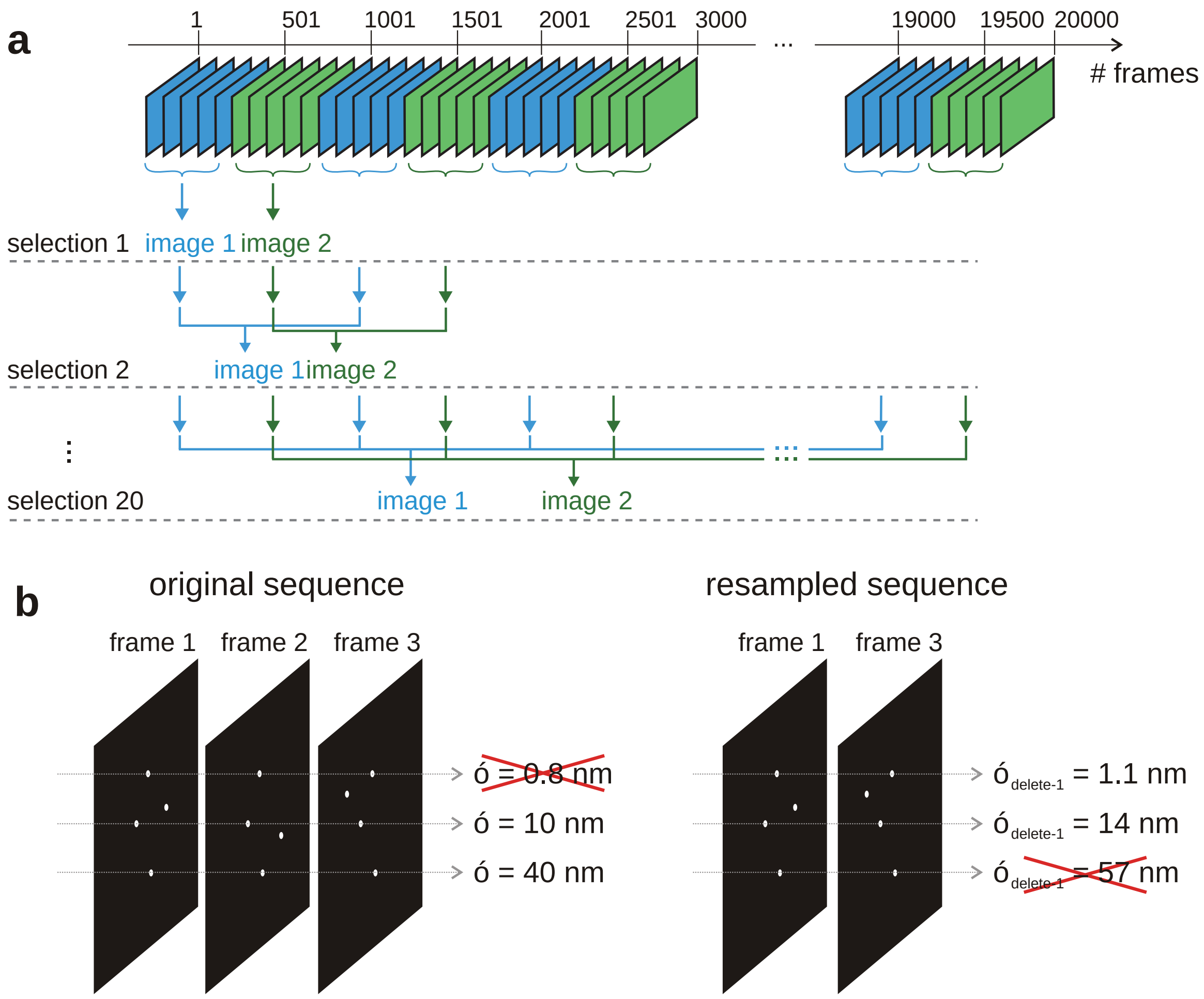

**Supplementary Figure 8:** Details of practical implementation of resolution and SNR metrics. (a) Illustration of the frame selection procedure for the sFRC calculation. (b) Illustration of the effect on the localization precision of leaving out a frame in the SNR calculation.

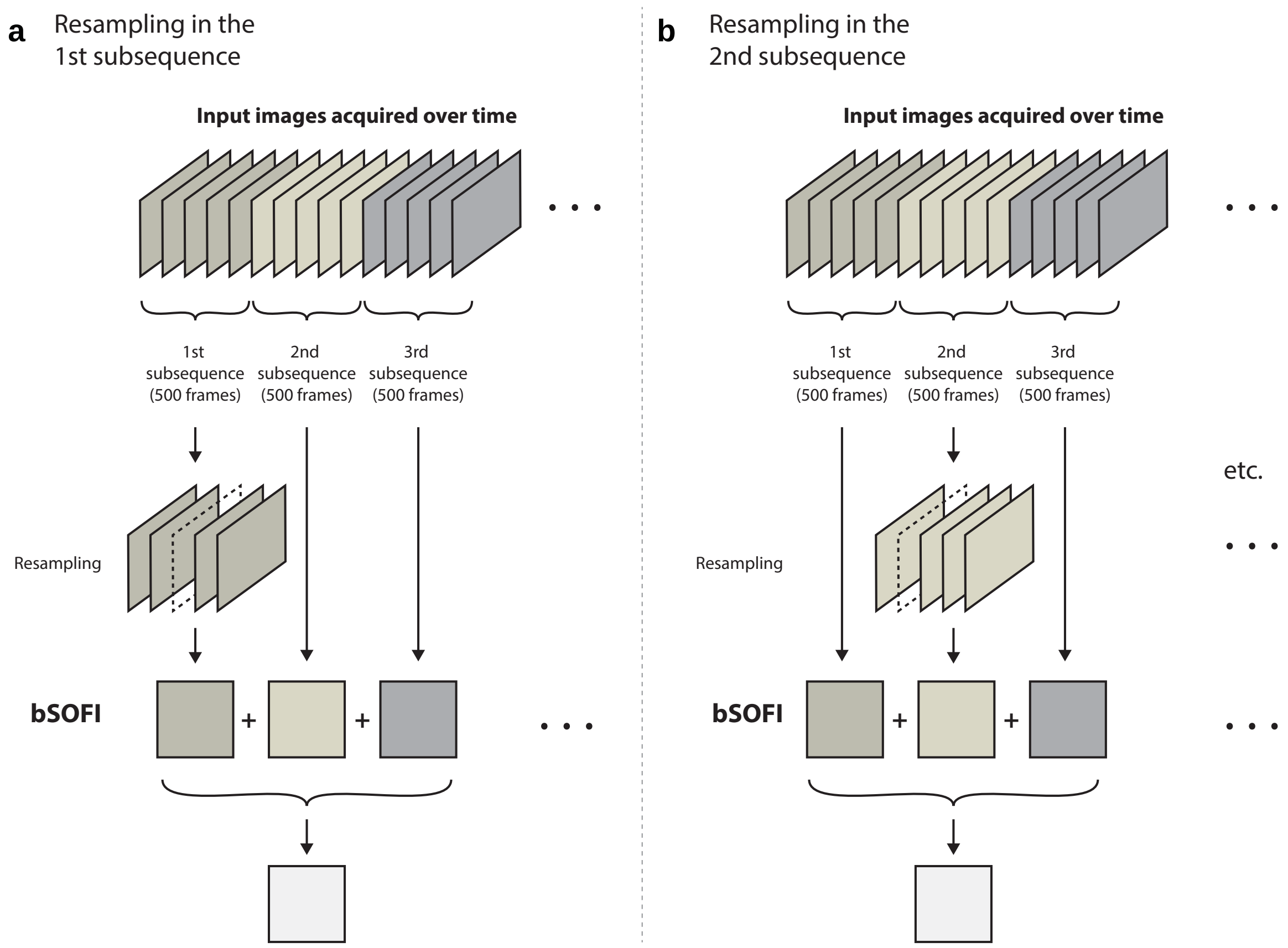

**Supplementary Figure 9:** Estimating SNR using jackknife resampling. (a) The input image sequence is divided into subsequences. In the first step, the resampling is performed within the first subsequence. Each time one frame of the first subsequence is left out, bSOFI image is calculated and summed up with the bSOFI images calculated from the remaining subsequences. (b) When all the resampling possibilities are evaluated in the first subsequence, the algorithm starts resampling the second subsequence.

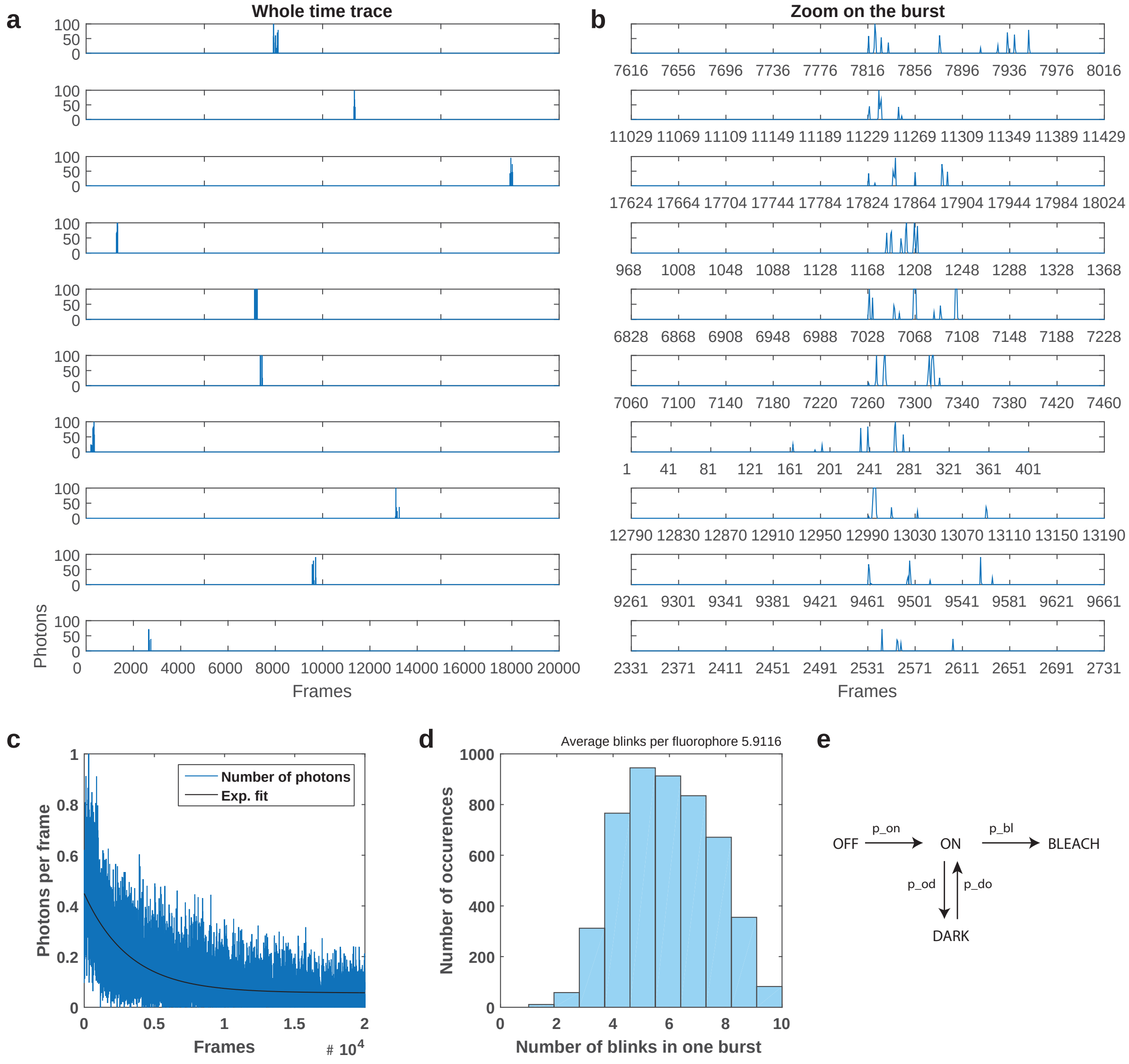

**Supplementary Figure 10:** Simulation of photophysics of fluorophores. (a) Time traces of the first 10 fluorophores are shown. Each fluorophore goes randomly into the on-state and during this "burst" quickly blinks several times i.e. switches between the bright and dark state. (b) The zoom shows these blinking events in detail. The frequency and duration of these blinks is modeled according to measurements of mEos2 photokinetics measured in [Durisic 2014, Nat. Met. Paper]. (c) Blinking statistics of simulated fluorophores. Number of photons as a function of frames. (d) Number of blinking events in on "burst". During the "burst" each fluorophore blinks several times (a random number in the range 2-10). (e) A schematic drawing of the four state photophysics model.

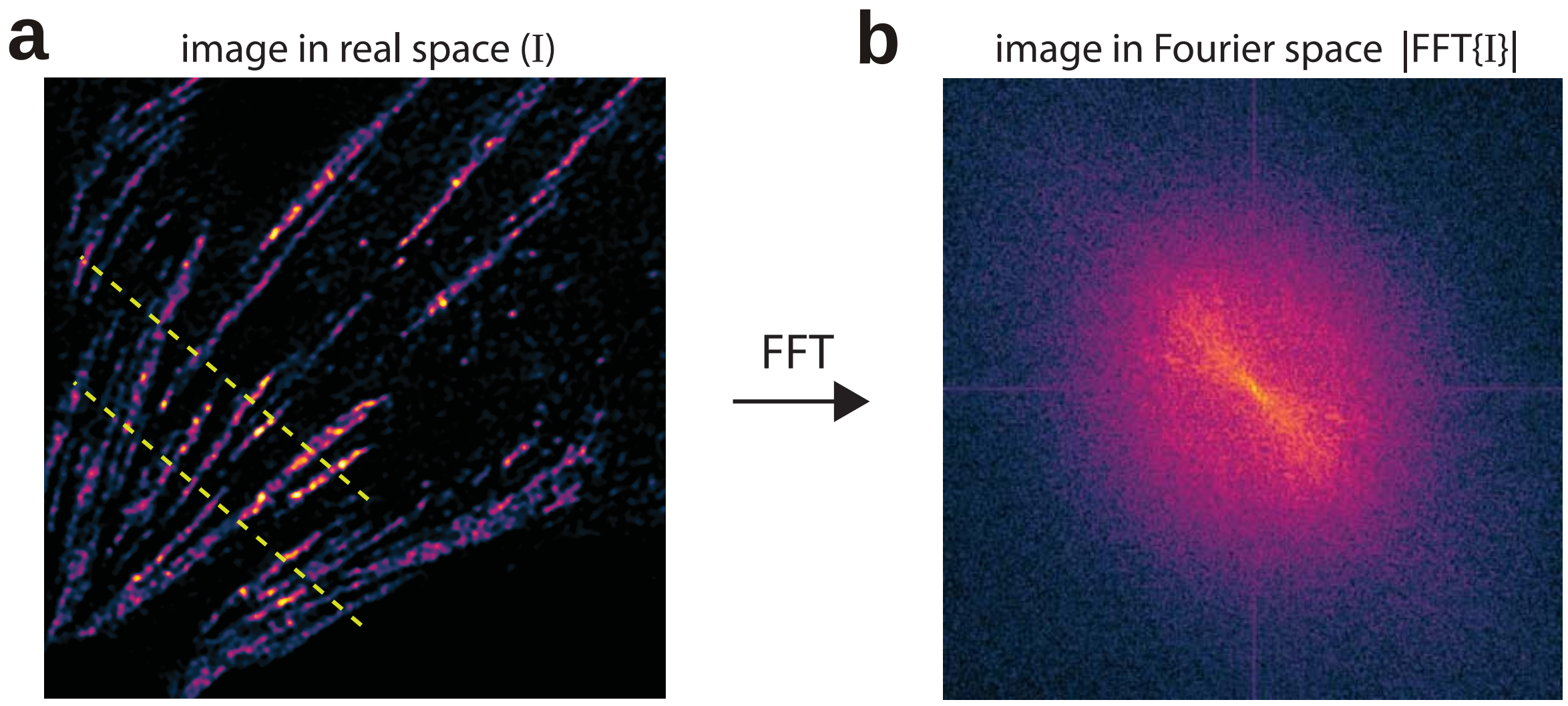

**Supplementary Figure 11:** (a) 4th order bSOFI image of a fixed MEF expressing paxillin labeled with psCFP2. The high spatial frequency changes of intensity appears mostly in one direction (marked by the yellow line). (b) Most of the high frequency content in Fourier space appears along the same direction. The color map "morgenstemning" was applied [10].

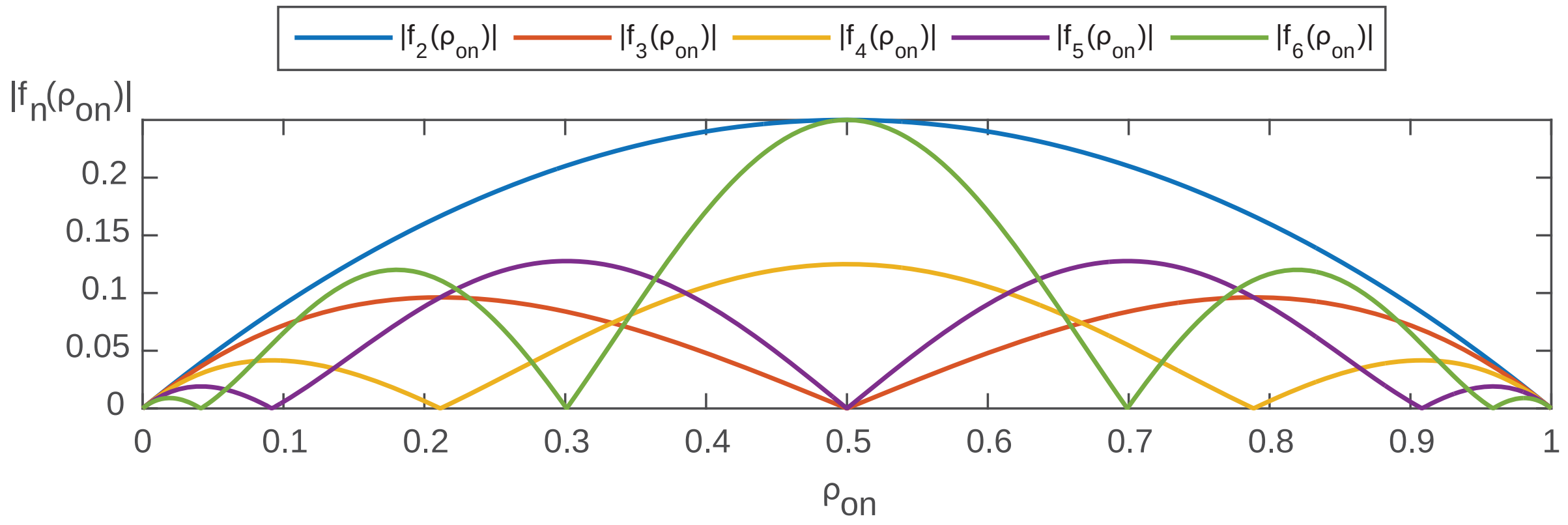

**Supplementary Figure 12:** On-time ratio polynomial of $2^{nd}$ to $6^{th}$ order as a function of the on-time ratio.

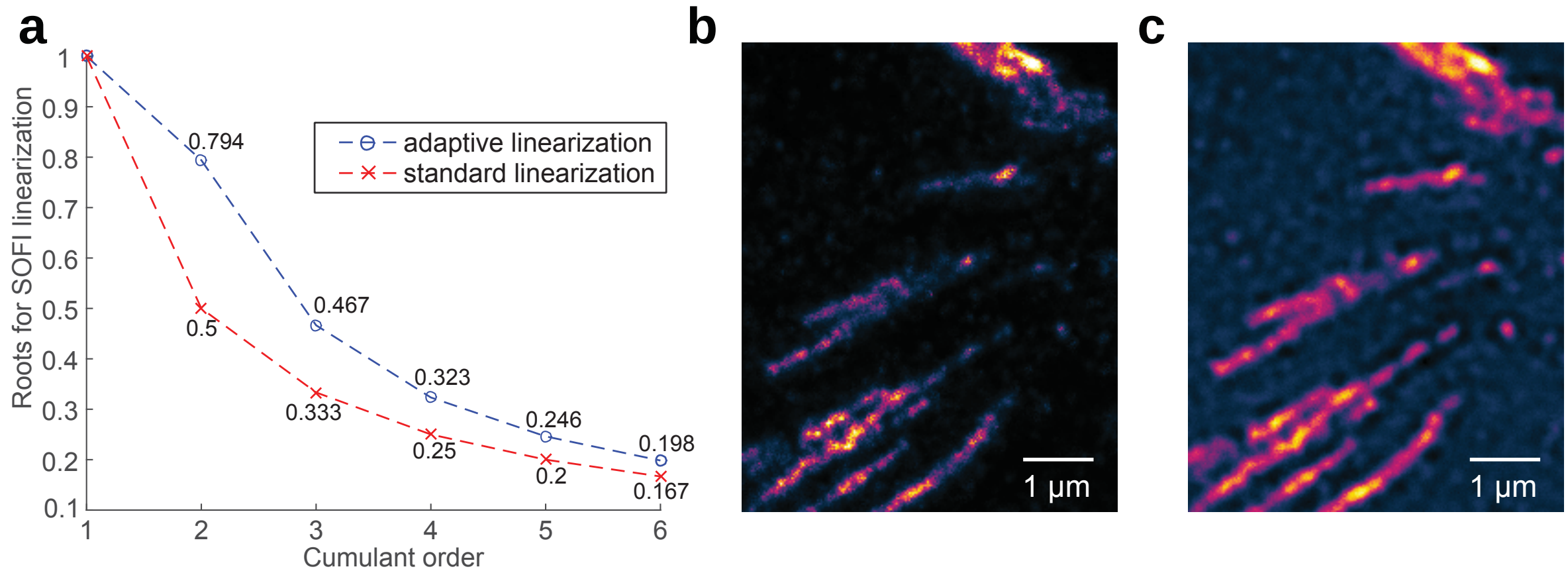

**Supplementary Figure 13:** Enhanced bSOFI. Images of fixed MEFs expressing paxillin labeled with mEos2 obtained from a raw image sequence of 20,000 frames. (a) Roots for SOFI standard and adaptive linearization. (b) 4th order bSOFI using a novel linearization (sFRC=134 nm). (c) 4th order bSOFI using standard linearization (sFRC=166 nm). Dynamic range is reduced too much which leads to lower SNR and deconvolution artifacts in the low SNR background regions.

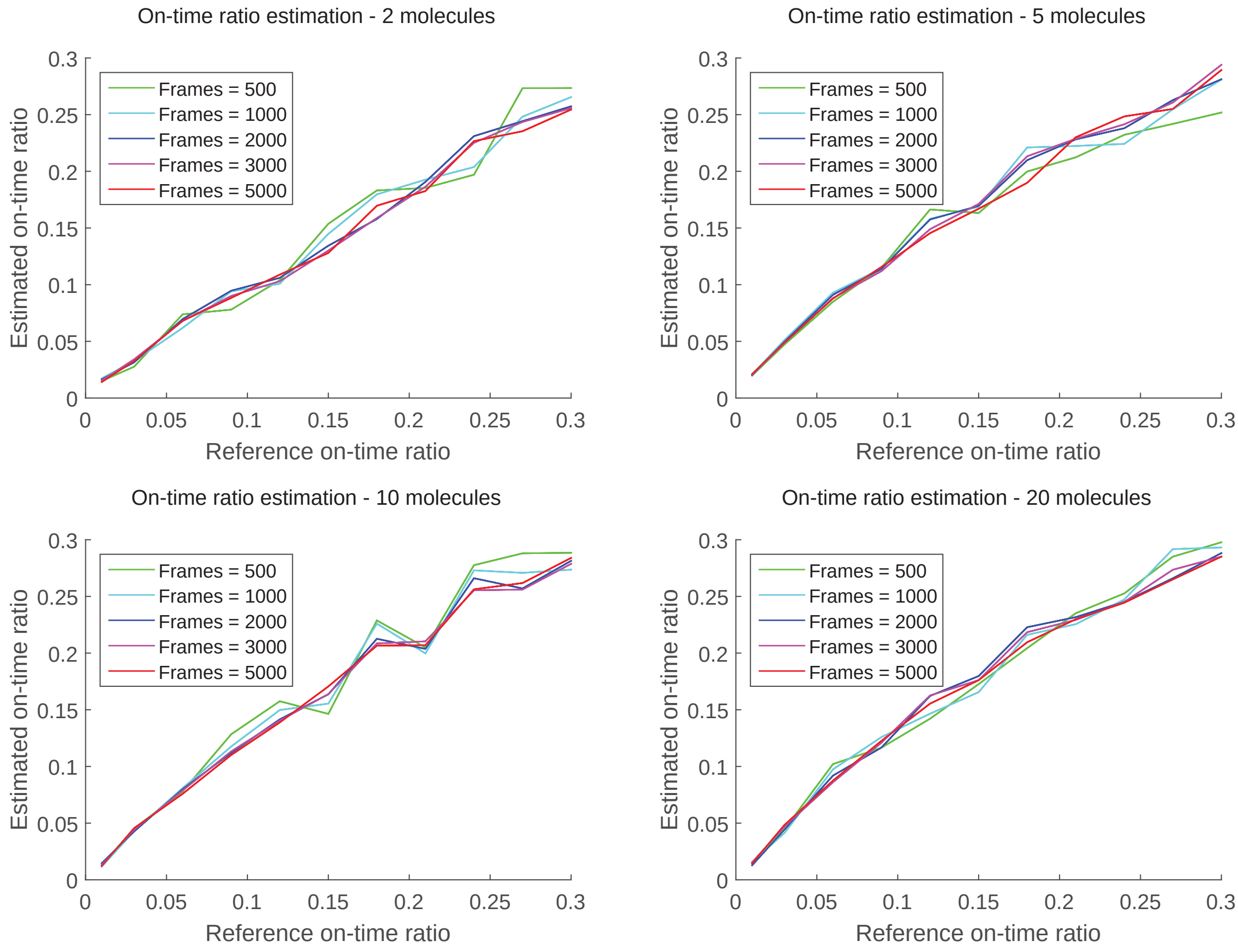

**Supplementary Figure 14:** On-time ratio estimation, tested with simulated input image stacks with a varying number of frames.